\documentclass[journal,twocolumn]{IEEEtran}
\usepackage{amsmath,amssymb}
\usepackage{graphicx}
\usepackage{cite}
\usepackage{booktabs}
\usepackage{bm}
\usepackage{mathtools}
\usepackage[hidelinks]{hyperref}
\usepackage{xcolor}
\usepackage[utf8]{inputenc}
\usepackage[T1]{fontenc}

\begin{document}
\title{To Stack or Not to Stack? Jacobian Conditioning and the
Benefits of Depth in Metasurface-Based Linear Transformation Synthesis}

\author{Andrea Abrardo,~\IEEEmembership{Senior Member,~IEEE,}
        and Andrea Garulli,~\IEEEmembership{Fellow,~IEEE}
\thanks{Andrea Abrardo and Andrea Garulli are with the
Department of Information Engineering and Mathematics (DIISM),
University of Siena, 53100 Siena, Italy.}
\thanks{Andrea Abrardo is also with the Consorzio Nazionale
Interuniversitario per le Telecomunicazioni (CNIT), Italy.}
}

\maketitle

\begin{abstract}
This paper investigates when and why stacking programmable metasurfaces into Stacked Intelligent Metasurfaces (SIMs) improves linear transformation synthesis at a fixed phase-control budget. The analysis shows that algebraic Jacobian rank alone is insufficient to characterize practical synthesis, which also depends on the strength and balance of the sensitivity directions generated by the available controls. A layer-resolved analysis of the cascade model explains how
inter-layer propagation can generate complementary sensitivity
directions. The relevance of this mechanism is then assessed
using an S-parameter multiport model that includes electromagnetic
coupling and multiple scattering.
Numerical results in near- and far-field sensing scenarios show that distributing the same control resources across multiple layers can strengthen weak transformation directions, improve Jacobian conditioning, and substantially enhance synthesis accuracy. These findings explain why increasing depth can offer benefits that are not obtained by simply enlarging a single programmable surface, and identify layer complementarity and Jacobian conditioning as useful criteria for SIM architecture design.
\end{abstract}

~\\
%%%%%%%%%%%%%%%%%%%%%%%%%%%%%%%%%%%%%%%%%%%%%%%%%%%%%%%%%%%%%%%%%%%%%%%%%%%
\section{Introduction}
\label{sec:introduction}
%%%%%%%%%%%%%%%%%%%%%%%%%%%%%%%%%%%%%%%%%%%%%%%%%%%%%%%%%%%%%%%%%%%%%%%%%%%
Future wireless systems increasingly combine communication, sensing,
localization, and data-driven intelligence
\cite{GonzalezPrelcic2024ISAC}. The associated processing demands and
constraints on latency and energy consumption motivate performing part of
the signal processing directly in the physical domain. Wave-domain
computing exploits propagation, interference, and scattering to implement
operations with intrinsic parallelism and potentially reduced digital
processing requirements
\cite{ZangenehNejad2021AnalogComputing,Qian2025IntelligentMetasurfaces}.

This approach has been extensively investigated in optics, where
diffractive networks implement input--output mappings by alternating
free-space propagation and programmable modulation
\cite{lin2018all,KulceCapacity2021,KulceArbitrary2021}.
At radio frequencies, stacked intelligent metasurfaces (SIMs) adopt a
related architecture: multiple programmable transmissive surfaces shape
and mix electromagnetic fields across successive propagation stages.
SIMs have been investigated for holographic MIMO, multiuser processing,
near-field focusing, integrated sensing and communications,
direction-of-arrival estimation, and localization
\cite{DiRenzoSIMSurvey,10158690,10922857,
DBLP:journals/ojcs/HassanARDY24,DBLP:conf/vtc/JiaALGRDY24,
DBLP:journals/wcl/NiuAPGCD24,An_2024_ref4,abrardo_Eusipco}.
These applications share a common synthesis problem: configuring the
metasurfaces to implement a prescribed linear transformation between
incident-field and output spaces. In sensing, for example, the desired
transformation may associate different arrival angles or source positions
with distinct output ports.

This perspective raises the architectural question addressed here:
\emph{when and why does stacking metasurfaces help if the total number of
programmable controls is unchanged?} The same phase budget can be
concentrated on one wide surface or distributed across several narrower
layers. Equal control counts, however, need not provide equally useful
transformation directions. Their strength and diversity also depend on
illumination, propagation-induced mixing, and output observations.
Even a full-rank Jacobian can contain weak singular directions that make a
prescribed transformation difficult to synthesize through iterative
optimization. The practical value of stacking therefore depends on how
it changes the conditioning of the phase-to-transformation map, beyond
simply increasing or preserving the number of accessible directions.

Most SIM studies use an ideal unilateral cascade of propagation matrices
and diagonal phase controls. This model makes the action of successive
layers explicit, but omits electromagnetic interactions that can be
significant in compact RF structures. Complete multiport formulations
account for mutual coupling, repeated scattering, and the port and cell
characteristics \cite{DR1,DR2,Abrardo24RisOpt,Abra_SIM1,PettaniceSIM2026}.
We use both descriptions: the cascade provides an interpretable mechanism
for the width--depth tradeoff, while the complete S-parameter model permits
its implications for conditioning and synthesis to be examined with these
interactions retained. This connects the architectural question to the
ability of optimization algorithms to realize useful sensing functions
at a fixed phase-control budget.

%%%%%%%%%%%%%%%%%%%%%%%%%%%%%%%%%%%%%%%%%%%%%%%%%%%%%%%%%%%%%%%%%%%%%%%%%%%
\subsection{Related Works and Motivation}
%%%%%%%%%%%%%%%%%%%%%%%%%%%%%%%%%%%%%%%%%%%%%%%%%%%%%%%%%%%%%%%%%%%%%%%%%%%
The capacity of diffractive processors to realize linear transformations,
and the effect of distributing a fixed number of trainable features across
layers, have been studied in optics
\cite{KulceCapacity2021,KulceArbitrary2021}. For reconfigurable electromagnetic
devices, \cite{Dardari2026Controllability} establishes sufficient conditions for controllability based on the rank of the real Jacobian of the parameter-to-response map. 
This provides a foundation for assessing realizability in principle.
Our focus is the practical synthesis problem: how the conditioning of that
map affects the ability of iterative algorithms to approach a prescribed
transformation. Rank alone does not quantify the phase changes needed to
correct a transformation error or the progress attainable along weak
sensitivity directions.

The relevance of the singular-value spectrum beyond algebraic rank is
already explicit in the backscatter MIMO analysis of
\cite{delHougne2026EEMDOF}. That work uses the participation number of a
multiport-derived Jacobian to quantify effective electromagnetic degrees of
freedom and examines their dependence on coupling and coherent illumination.
Here, we connect transformation sensitivities to the local inverse synthesis
problem: weak singular directions can require large Gauss--Newton
corrections and yield small gradient components, depending on the alignment
of the residual with those directions. This connection allows the spectrum
to be interpreted in terms of optimization behavior and used to investigate
the allocation of a fixed phase budget between width and depth.

%Electromagnetically consistent upper bounds on target-operator fidelity under hardware constraints are derived in \cite{delHougne2026Bounds}. These bounds provide an algorithm-independent benchmark for assessing how closely a synthesized operator approaches the best achievable fidelity. Here, we investigate how iterative optimization can reach an accurate realization of a prescribed transformation, using a normalized mean-square error criterion with optimal complex scaling. A small relative error directly certifies accurate reproduction of the target up to that scaling, without requiring a separate bound on the globally optimal performance. Our focus is therefore on how Jacobian conditioning facilitates or hinders reaching such low-error solutions, and on why distributing a fixed phase-control budget across multiple layers can improve this process.
Within this framework, stacking can be useful because inter-layer
propagation changes the fields on which each set of controls acts, producing
sensitivity directions that complement those available on a single surface.
We examine this mechanism and characterize the
resulting spectra through a relative-threshold effective rank: the number
of Jacobian singular values that remain above a prescribed fraction of the
largest one. In the sensing case studies, this measure closely distinguishes
configurations that attain small synthesis errors with Gauss--Newton and gradient descent optimization from those that do not. It thereby provides a
concrete diagnostic linking the distribution of control sensitivities to
practical synthesis performance.

%%%%%%%%%%%%%%%%%%%%%%%%%%%%%%%%%%%%%%%%%%%%%%%%%%%%%%%%%%%%%%%%%%%%%%%%%%%
\subsection{Main Contributions}
%%%%%%%%%%%%%%%%%%%%%%%%%%%%%%%%%%%%%%%%%%%%%%%%%%%%%%%%%%%%%%%%%%%%%%%%%%%
The main contributions are as follows.
\begin{itemize}
\item We formulate transformation synthesis locally as an inverse problem
and relate the Jacobian singular values and residual alignment to Gauss--Newton corrections and gradient-based progress.  This identifies how
poor conditioning can hinder practical synthesis despite full local rank.

\item We derive a layer-resolved Jacobian frame interpretation for the
unilateral cascade. It explains how lateral enlargement can produce
geometric saturation and how depth can supply complementary directions,
which can significantly improve the Jacobian conditioning.

\item We evaluate exact transformation sensitivities in the complete
multiport model and compare wide and deep SIMs at equal phase budget in
near-field position-discrimination and far-field angular-discrimination
sensing tasks, using both Gauss--Newton with line search and gradient-based optimization. The relative-threshold effective rank of the Jacobian closely tracks the synthesis accuracy attained in the considered experiments. These results support the cascade-based interpretation: at fixed total phase budget, distributing controls across several layers yield better-balanced sensitivities than concentrating them on a wider output-facing surface.  

\end{itemize}

The paper is organized as follows. Section~\ref{sec:S_multiport_model} introduces the multiport model and the synthesis problem. Section~\ref{sec:optimization} relates local optimization to the Jacobian spectrum. Section~\ref{sec:cascade_jacobian_interpretation} develops the Jacobian structure of the cascade model and provides geometric interpretation for the role of depth. Section~\ref{sec:multiport_jacobian} derives the exact multiport sensitivities, and Section~\ref{sec:results} presents the numerical comparison of wide and deep architectures.

\textit{Notation}: Bold lowercase and uppercase letters denote vectors and matrices. The superscripts $T$, $H$, and $\dagger$ denote transpose, conjugate transpose, and Moore--Penrose pseudoinverse, respectively. The operator $\operatorname{vec}(\cdot)$ stacks matrix columns, while $\Re\{\cdot\}$ and $\Im\{\cdot\}$ extract real and imaginary parts. The symbols $\otimes$, $\|\cdot\|_2$, and $\|\cdot\|_F$ denote the Kronecker product, Euclidean norm, and Frobenius norm, respectively. The imaginary unit is $j$.

%%%%%%%%%%%%%%%%%%%%%%%%%%%%%%%%%%%%%%%%%%%%%%%%%%%%%%%%%%%%%%%%%%%%%%%%%%%
%%%%%%%%%%%%%%%%%%%%%%%%%%%%%%%%%%%%%%%%%%%%%%%%%%%%%%%%%%%%%%%%%%%%%%%%%%%

\section{System Model and Problem Formulation}
\label{sec:S_multiport_model}
%%%%%%%%%%%%%%%%%%%%%%%%%%%%%%%%%%%%%%%%%%%%%%%%%%%%%%%%%%%%%%%%%%%%%%%%%%%

We adopt the S-parameter multiport formulation developed for SIMs in
\cite{PettaniceSIM2026}, which is equivalent, under the corresponding
network transformations, to the impedance-based description in
\cite{Abra_SIM1}. This representation describes both electromagnetic
coupling and tunable multiport cells within a common scattering framework.

The model comprises $N_T$ transmitter ports, $N_E$ programmable
electromagnetic ports associated with the metasurface cells, and $N_R$
receiver ports. The fixed electromagnetic structure is described by the
scattering blocks $\mathbf S_{ET}$, $\mathbf S_{EE}$,
$\mathbf S_{RE}$, and $\mathbf S_{RT}$, accounting, respectively, for
transmitter-to-programmable-port coupling, coupling among programmable
ports, programmable-port-to-receiver coupling, and direct
transmitter-to-receiver coupling. The programmable circuitry is represented
by its scattering matrix
$\boldsymbol{\Gamma}(\boldsymbol{\phi})\in\mathbb C^{N_E\times N_E}$,
where $\boldsymbol{\phi}=[\phi_1,\ldots,\phi_P]^T\in\mathbb R^P$
collects the independently controlled phase parameters.

Under matched transmitter and receiver terminations, the end-to-end
transfer matrix mapping incident power-wave vectors at the transmitter
ports to outgoing power-wave vectors at the receiver ports is
\cite{PettaniceSIM2026}
\begin{equation}
\mathbf H(\boldsymbol{\phi})
=
\mathbf S_{RT}
+
\mathbf S_{RE}\boldsymbol{\Gamma}(\boldsymbol{\phi})
\mathbf F(\boldsymbol{\phi})\mathbf S_{ET},
\label{eq:H_S}
\end{equation}
where
\begin{equation}
\mathbf F(\boldsymbol{\phi})
=
\left(
\mathbf I-\mathbf S_{EE}\boldsymbol{\Gamma}(\boldsymbol{\phi})
\right)^{-1}.
\label{eq:F_S}
\end{equation}
The first term in \eqref{eq:H_S} describes the direct path, while the
second includes all contributions mediated by the programmable structure.
In particular, $\mathbf S_{EE}$ and the inverse in \eqref{eq:F_S}
retain inter-element and inter-layer coupling and repeated scattering.
Consequently, changing one programmable phase can modify the complete
internal field and hence the overall transfer matrix.

The structure of $\boldsymbol{\Gamma}(\boldsymbol{\phi})$ depends on
the programmable architecture. For a conventional single-layer RIS with
independently terminated, ideal lossless phase-only ports, it is a diagonal matrix,
with reflection coefficients $e^{j\phi_p}$ on the diagonal.
For the multilayer transmissive SIM considered here, each programmable
cell is a tunable two-port network connecting electromagnetic ports
associated with adjacent regions of the stack. Thus,
$\boldsymbol{\Gamma}(\boldsymbol{\phi})$ is block diagonal across the
cells, with each block controlled by one phase parameter.
A SIM with $Q$ layers and $N$ cells per layer therefore has $P=QN$
independent phase controls. The cell model can include insertion loss,
return loss, and amplitude--phase coupling, with an ideal reciprocal
phase-only cell as a special case. The block structure and its assembly
are detailed in \cite{PettaniceSIM2026}.

Beyond-diagonal RISs pursue a complementary architectural direction:
programmable interconnections between distinct elements introduce
controllable coupling within a single physical layer
\cite{Li_2023_ref1,Nerini_Clerckx_SIM}. Depending on the network topology
and parameterization, a single control can then affect scattering among
multiple ports, and its sensitivity can depend on the other control
settings. This differs from the cell-wise block structure considered
here, where each control acts on one two-port cell and electromagnetic
coupling between cells is represented by $\mathbf S_{EE}$.
Beyond-diagonal architectures therefore offer an alternative to increasing
depth; their expressivity analysis is outside the scope of this work,
which focuses on distributing independently controlled cells across
multiple layers at a fixed total phase budget.

To describe simultaneously the response to multiple excitations, consider
an arbitrary collection of $I$ independent excitations
\begin{equation}
\mathbf X
=
\begin{bmatrix}
\mathbf x_1 & \cdots & \mathbf x_I
\end{bmatrix}
\in\mathbb C^{N_T\times I}.
\label{eq:X_multiport}
\end{equation}
The vectors \(\mathbf x_i\) may represent, depending on the considered
scenario, different transmitter excitations, incident wavefronts, spatial
modes, or fields generated by sources located at different positions.
We further introduce
$\mathbf C\in\mathbb C^{M\times N_R}$ to account for any additional linear
mapping between the $N_R$ receiver ports represented by the multiport model
and the $M$ output observations of interest. This mapping may represent,
for example, propagation outside the SIM toward spatially separated
receivers, port selection, or linear combining at the receiver. The
corresponding output matrix is
\begin{equation}
\mathbf Y(\boldsymbol{\phi})
=
\mathbf C\mathbf H(\boldsymbol{\phi})\mathbf X
\in\mathbb C^{M\times I}.
\label{eq:Y_multiport}
\end{equation}

Defining
\begin{equation}
\mathbf Y_d=\mathbf C\mathbf S_{RT}\mathbf X,\qquad
\mathbf A=\mathbf C\mathbf S_{RE},\qquad
\mathbf B=\mathbf S_{ET}\mathbf X,
\end{equation}
we obtain the compact representation
\begin{equation}
\mathbf Y(\boldsymbol{\phi})
=
\mathbf Y_d+
\mathbf A\boldsymbol{\Gamma}(\boldsymbol{\phi})
\left(
\mathbf I-\mathbf S_{EE}\boldsymbol{\Gamma}(\boldsymbol{\phi})
\right)^{-1}
\mathbf B.
\label{eq:Y_compact_S}
\end{equation}
The direct term $\mathbf Y_d$ is independent of the programmable phases.
In the following derivations we omit this fixed contribution from the
controlled transformation. %If the direct path is retained and known,
%it is added back when evaluating the output and residual; all phase
%Jacobians are unchanged. For an unscaled target, this is equivalently
%handled by subtracting $\mathbf Y_d$ from the target. When the complex
%output scaling is fitted, that fit must still use the total output,
%including the direct term. Thus omitting $\mathbf Y_d$ simplifies the
%notation without changing the sensitivity derivations, but does not
%assert that an arbitrary direct path leaves synthesis accuracy unchanged.

%%%%%%%%%%%%%%%%%%%%%%%%%%%%%%%%%%%%%%%%%%%%%%%%%%%%%%%%%%%%%%%%%%%%%%%%%%%
\subsection{Relation to the Unilateral Cascade}
\label{sec:relation_cascade}
%%%%%%%%%%%%%%%%%%%%%%%%%%%%%%%%%%%%%%%%%%%%%%%%%%%%%%%%%%%%%%%%%%%%%%%%%%%

The relation between the complete multiport representation and the
conventional unilateral cascade has been investigated in detail in
\cite{Abra_SIM1}. The latter can be recovered as an idealized representation
when the relevant electromagnetic interactions are restricted to
unilateral propagation between consecutive layers and the programmable
cells are modeled as ideal phase-only elements.

Let $\mathbf S_{\rm in}$ denote the mapping from the transmitter
excitations to the field incident on the first programmable layer, and let
$\mathbf S_{\rm out}$ denote the mapping from the last programmable layer
to the $M$ output observations. Under ideal phase-only cells, negligible
undesired mutual coupling and reflections, and unilateral inter-layer
propagation, the input--output response assumes the cascaded form
\begin{equation}
\mathbf Y_{\rm cas}(\boldsymbol{\phi})
=
\mathbf S_{\rm out}
\mathbf D_Q\mathbf S_Q\mathbf D_{Q-1}\cdots
\mathbf S_2\mathbf D_1
\mathbf S_{\rm in}\mathbf X,
\label{eq:Y_cascade_limit}
\end{equation}
where
\begin{equation}
\mathbf D_q
=
\operatorname{diag}
\left(
e^{j\phi_{q,1}},\ldots,e^{j\phi_{q,N}}
\right)
\end{equation}
collects the $N$ programmable phase shifts of layer $q$, while $\mathbf S_q$
describes unilateral propagation between consecutive programmable layers.

{In this representation, $\mathbf S_{\rm in}\mathbf X$ plays the role of
the excitation delivered to the programmable structure, analogous to
$\mathbf S_{ET}\mathbf X$ in the complete multiport formulation, whereas
$\mathbf S_{\rm out}$ represents the mapping from the last programmable
layer to the output observations, corresponding to the role played by
$\mathbf C\mathbf S_{RE}$ in the complete model. The cascaded formulation
thus absorbs the propagation stages before the first and after the last
programmable layer into $\mathbf S_{\rm in}$ and $\mathbf S_{\rm out}$,
respectively.} 

The cascade in \eqref{eq:Y_cascade_limit} provides a substantially simpler
and analytically tractable representation of the SIM. However, unlike
\eqref{eq:Y_compact_S}, it does not retain the global reciprocal
electromagnetic interactions represented by $\mathbf S_{EE}$ and by the
matrix inverse in the complete multiport model. In the following, we exploit
this idealized cascade 
to analyze analytically some key expressivity mechanisms.
The insights obtained from this model will subsequently be assessed using the
complete multiport representation.

%%%%%%%%%%%%%%%%%%%%%%%%%%%%%%%%%%%%%%%%%%%%%%%%%%%%%%%%%%%%%%%%%%%%%%%%%%%
\subsection{Linear Transformation Synthesis Problem}
\label{sec:linear_transformation_synthesis}
%%%%%%%%%%%%%%%%%%%%%%%%%%%%%%%%%%%%%%%%%%%%%%%%%%%%%%%%%%%%%%%%%%%%%%%%%%%

Let $\mathbf X$ and $\mathbf Y(\boldsymbol{\phi})$ denote the input
and output matrices introduced in \eqref{eq:X_multiport} and
\eqref{eq:Y_multiport}. We assume $\operatorname{rank}(\mathbf X)=I$,
so that the selected excitations form a basis for the input subspace
of interest. A prescribed linear operation on this subspace is specified
by the nonzero target matrix
\begin{equation}
\mathbf T_0
=
\begin{bmatrix}
\mathbf t_1 & \cdots & \mathbf t_I
\end{bmatrix}
\in\mathbb C^{M\times I},
\label{eq:target_transformation}
\end{equation}
where $\mathbf t_i$ is the desired response to $\mathbf x_i$.

We seek to reproduce the relative complex structure of $\mathbf T_0$
up to a global complex factor, since the overall gain is constrained
by propagation, losses, and the passive architecture. The synthesis
objective is thus to find $\boldsymbol{\phi}$ such that
\begin{equation}
\beta\mathbf Y(\boldsymbol{\phi})
\simeq
\mathbf T_0,
\label{eq:scaled_synthesis}
\end{equation}
for some $\beta\in\mathbb C$.
The synthesis mismatch is measured by the normalized squared error
\begin{equation}
\epsilon(\boldsymbol{\phi})
=
\min_{\beta\in\mathbb C}
\frac{
\left\|
\beta\mathbf Y(\boldsymbol{\phi})-\mathbf T_0
\right\|_F^2
}{
\|\mathbf T_0\|_F^2
}.
\label{eq:SI_cost}
\end{equation}
For $\mathbf Y(\boldsymbol{\phi})\neq\mathbf 0$, the minimizing
coefficient is
\begin{equation}
\beta_\star(\boldsymbol{\phi})
=
\frac{
\operatorname{vec}\!\left(\mathbf Y(\boldsymbol{\phi})\right)^H
\operatorname{vec}(\mathbf T_0)
}{
\|\mathbf Y(\boldsymbol{\phi})\|_F^2
}.
\label{eq:optimal_beta}
\end{equation}

%%%%%%%%%%%%%%%%%%%%%%%%%%%%%%%%%%%%%%%%%%%%%%%%%%%%%%%%%%%%%%%%%%%%%%%%%%%
\section{Local Inverse Synthesis and Optimization}
\label{sec:optimization}
%%%%%%%%%%%%%%%%%%%%%%%%%%%%%%%%%%%%%%%%%%%%%%%%%%%%%%%%%%%%%%%%%%%%%%%%%%%

{We now approach the transformation-synthesis problem
introduced in Section~\ref{sec:linear_transformation_synthesis} by performing a sequence of local updates of the phase configuration $\boldsymbol{\phi}$.}
At a given $\boldsymbol{\phi}$, the optimal complex scaling
coefficient $\beta_\star$ is first evaluated according to
\eqref{eq:optimal_beta} and is then kept fixed while constructing the local
phase update. After the phases are updated, $\beta_\star$ is recomputed at
the new configuration.

Let
\begin{equation}
\boldsymbol{\tilde{y}}(\boldsymbol{\phi})
=
\beta_\star
\operatorname{vec}\!\left(\mathbf Y(\boldsymbol{\phi})\right),
\qquad
\boldsymbol{\tilde{t}}
=
\operatorname{vec}(\mathbf T_0).
\label{eq:local_complex_vectors}
\end{equation}
Since the programmable phases are real-valued whereas the transformation is
complex-valued, we introduce the corresponding real vectors
\begin{equation}
\mathbf y(\boldsymbol{\phi})
=
\begin{bmatrix}
\Re\{\boldsymbol{\tilde{y}}(\boldsymbol{\phi})\}\\
\Im\{\boldsymbol{\tilde{y}}(\boldsymbol{\phi})\}
\end{bmatrix},
\qquad
\mathbf y_{\rm t}
=
\begin{bmatrix}
\Re\{\boldsymbol{\tilde{t}}\}\\
\Im\{\boldsymbol{\tilde{t}}\}
\end{bmatrix}.
\label{eq:real_transformation_vectors}
\end{equation}
The real Jacobian associated with the transformation is therefore
\begin{equation}
\mathbf J(\boldsymbol{\phi})
=
\frac{\partial\mathbf y(\boldsymbol{\phi})}
{\partial\boldsymbol{\phi}}
\in\mathbb R^{2MI\times P},
\label{eq:real_synthesis_jacobian}
\end{equation}
where $\beta_\star$ is treated as constant in the differentiation.

For a sufficiently small phase perturbation
$\delta\boldsymbol{\phi}$, the transformation generated by the SIM is
locally approximated as
\begin{equation}
\mathbf y^{\rm upd}
\simeq
\mathbf y
+
\mathbf J\delta\boldsymbol{\phi}.
\label{eq:local_transformation_update}
\end{equation}
Defining the current transformation error as
$
\mathbf e
=
\mathbf y_{\rm t}
-
\mathbf y,
%\label{eq:local_error}
$
the phase perturbation required to compensate for the error within the
first-order approximation should satisfy
\begin{equation}
\mathbf J\delta\boldsymbol{\phi}
\simeq
\mathbf e.
\label{eq:local_inverse_problem}
\end{equation}

Equation~\eqref{eq:local_inverse_problem} is the local inverse synthesis
problem underlying the optimization of the programmable phases. Its
properties depend not only on the rank of $\mathbf J$, which determines
the number of independent transformation directions that can be generated
locally, but also on its singular-value spectrum, which determines how
strongly those directions respond to variations of the programmable phases.

%%%%%%%%%%%%%%%%%%%%%%%%%%%%%%%%%%%%%%%%%%%%%%%%%%%%%%%%%%%%%%%%%%%%%%%%%%%
%\subsection{Jacobian Spectrum and Local Inverse Optimization}
%\label{sec:jacobian_spectrum}
%%%%%%%%%%%%%%%%%%%%%%%%%%%%%%%%%%%%%%%%%%%%%%%%%%%%%%%%%%%%%%%%%%%%%%%%%%%

In the following, we recall some properties of local inverse optimization which depend on the Jacobian spectrum.
Let $D=2MI$ and assume
\begin{equation}
P\geq D,
\qquad
\operatorname{rank}(\mathbf J)=D,
\label{eq:full_row_rank_regime}
\end{equation}
which represents the case of primary interest in this work. Full row rank of the
transformation Jacobian implies that all transformation directions are locally accessible. The
question addressed here is not whether these directions exist,
but how effectively they can be exploited by an optimization algorithm.

Consider the singular-value decomposition
$
\mathbf J
=
\mathbf U\boldsymbol{\Sigma}\mathbf V^T,
%\label{eq:J_SVD}
$
where
$
\sigma_1\geq\sigma_2\geq\cdots\geq\sigma_D>0,
$
are the singular values of $\mathbf J$. 
Since $\mathbf J$ has full row rank, its $D$ left singular vectors form an
orthonormal basis of the $D$-dimensional transformation space. The current
local error can therefore be represented as
\begin{equation}
\mathbf e
=
\mathbf U\mathbf c,
\qquad
\mathbf c
=
\mathbf U^T\mathbf e,
\label{eq:error_SVD_decomposition}
\end{equation}
where $c_i$ is the error component along the $i$th left singular direction.

%%%%%%%%%%%%%%%%%%%%%%%%%%%%%%%%%%%%%%%%%%%%%%%%%%%%%%%%%%%%%%%%%%%%%%%%%%%
\subsubsection{Gauss--Newton Interpretation}
%%%%%%%%%%%%%%%%%%%%%%%%%%%%%%%%%%%%%%%%%%%%%%%%%%%%%%%%%%%%%%%%%%%%%%%%%%%

Consider the classical Gauss--Newton correction obtained through
the Moore--Penrose pseudoinverse:
\begin{equation}
\delta\boldsymbol{\phi}_{\rm GN}
=
\mathbf J^{\dagger}\mathbf e
=
\sum_{i=1}^{D}
\frac{c_i}{\sigma_i}\mathbf v_i,
\label{eq:GN_pseudoinverse}
\end{equation}
where the full-row-rank assumption
ensures $\sigma_i>0$ for all $i$.
The squared norm of the correction is therefore
\begin{equation}
\|\delta\boldsymbol{\phi}_{\rm GN}\|_2^2
=
\sum_{i=1}^{D}\frac{|c_i|^2}{\sigma_i^2}.
\label{eq:GN_correction_norm}
\end{equation}
Thus, error components aligned with weak singular directions
require larger phase corrections than equally sized components
aligned with strong directions.

In the local linear model, the full Gauss--Newton step cancels
the error, i.e.,
$\mathbf e^{\rm upd} = \mathbf y_{\rm t} - \mathbf y^{\rm upd}
\simeq
\mathbf e-\mathbf J\delta\boldsymbol{\phi}_{\rm GN}
=
\mathbf 0$.
For the nonlinear transformation, however, this prediction is
reliable only when the phase correction is sufficiently small.
A large correction driven by a weak singular direction may leave
the region in which the linear approximation is accurate, making
step reduction necessary. For a scaled step
$\alpha\delta\boldsymbol{\phi}_{\rm GN}$, the predicted residual is
$
\mathbf e^{\rm upd}\simeq(1-\alpha)\mathbf e.
$
Consequently, reducing the step to accommodate weak inverse
directions also reduces the predicted progress along all other
directions.

The relative imbalance between these sensitivities is quantified by
the inverse condition number
\begin{equation}
\gamma
=
\frac{\sigma_{\min}(\mathbf J)}
     {\sigma_{\max}(\mathbf J)},
%\qquad
%\kappa_2(\mathbf J)=\frac{1}{\gamma}.
\label{eq:GN_relative_conditioning}
\end{equation}

For equal-magnitude residual components, the correction required
along the weakest singular direction is $1/\gamma$ times that
required along the strongest direction. A small value of $\gamma$
therefore indicates strongly unbalanced inverse sensitivities,
which can hinder practical Gauss--Newton synthesis despite
full local rank.

%%%%%%%%%%%%%%%%%%%%%%%%%%%%%%%%%%%%%%%%%%%%%%%%%%%%%%%%%%%%%%%%%%%%%%%%%%%
\subsubsection{Gradient-Descent Interpretation}
%%%%%%%%%%%%%%%%%%%%%%%%%%%%%%%%%%%%%%%%%%%%%%%%%%%%%%%%%%%%%%%%%%%%%%%%%%%

The Jacobian spectrum also determines the local behavior of
gradient descent. 
When minimizing
$\frac{1}{2}\|\mathbf e\|_2^2$, a gradient step with step size
$\mu>0$ is
\begin{equation}
\delta\boldsymbol{\phi}_{\rm GD}
=
\mu\mathbf J^T\mathbf e.
\label{eq:GD_update}
\end{equation}
Within the local linear approximation, the updated residual is
$
\mathbf e^{\rm upd}
\simeq
\left(\mathbf I-\mu\mathbf J\mathbf J^T\right)\mathbf e.
$
Therefore, its components along the left singular directions satisfy
$
c_i^{\rm upd}
\simeq
\left(1-\mu\sigma_i^2\right)c_i.
%\label{eq:GD_component_update}
$
For the local quadratic model, choosing
$\mu=\eta/\sigma_{\max}^2$ with $0<\eta<2$ ensures contraction
of every residual component under the full-row-rank assumption
\cite[Sec.~9.3]{BoydVandenberghe2004}, yielding:
\begin{equation}
c_i^{\rm upd}
\simeq
\left(1-\eta\frac{\sigma_i^2}{\sigma_{\max}^2}\right)c_i.
\label{eq:GD_relative_rate}
\end{equation}
In particular, along the weakest singular direction,
\begin{equation}
c_{\min}^{\rm upd}
\simeq
\left(1-\eta\gamma^2\right)c_{\min},
%\qquad
%\gamma=\frac{\sigma_{\min}}{\sigma_{\max}}.
\end{equation}
For a fixed $\eta$, this factor approaches one as $\gamma$
decreases, indicating increasingly slow correction of the
corresponding error component.

Thus, weak singular directions affect the two methods differently:
Gauss--Newton may require large phase corrections, whereas
gradient descent makes limited progress along those directions
when its step size is constrained by the strongest sensitivity.
In both cases, the practical impact depends on the residual
components aligned with the weak directions. These
considerations motivate examining the relative strength of
accessible transformation directions, rather than their number
alone.

Hereafter, we distinguish the scaled synthesis Jacobian $J$ in \eqref{eq:real_synthesis_jacobian} from the unscaled
transformation Jacobians
\begin{equation}
\mathbf J_c=\frac{\partial\operatorname{vec}(\mathbf Y)}{\partial\boldsymbol{\phi}},
\qquad
\mathbf J_R=\begin{bmatrix}\Re\{\mathbf J_c\}\\\Im\{\mathbf J_c\}\end{bmatrix}.
\label{eq:real_unscaled_jacobian}
\end{equation}
Thus, $\mathbf J\!=\![\Re\{\beta_\star\mathbf J_c\}^T,
\Im\{\beta_\star\mathbf J_c\}^T]^T$, and $\sigma_i(\mathbf J)\!=\!|\beta_\star|\,\sigma_i(\mathbf J_R)$. Hence, for $\beta_\star\neq0$,
$\mathbf J$ and $\mathbf J_R$ have the same rank, condition number and  normalized spectrum. Therefore, 
the subsequent architectural analysis will be carried out on $\mathbf J_R$.

%%%%%%%%%%%%%%%%%%%%%%%%%%%%%%%%%%%%%%%%%%%%%%%%%%%%%%%%%%%%%%%%%%%%%%%%%%%
\section{Cascaded Interpretation of the Jacobian Conditioning}
\label{sec:cascade_jacobian_interpretation}
%%%%%%%%%%%%%%%%%%%%%%%%%%%%%%%%%%%%%%%%%%%%%%%%%%%%%%%%%%%%%%%%%%%%%%%%%%%

The synthesis problem introduced above raises the question of how
effectively the physical phase controls can generate perturbations of the
desired input--output transformation. The number of controls alone does
not answer this question: their sensitivity directions may be strongly
correlated, yielding a poorly conditioned Jacobian even when its
algebraic rank is large. We examine this distinction through the ideal
unilateral cascade introduced in Section~\ref{sec:relation_cascade},
whose factorized structure exposes the contribution of each control.
The resulting interpretation will subsequently be assessed using the
complete multiport model.

Starting from \eqref{eq:Y_cascade_limit}, let
$\mathbf W^T=\mathbf S_{\rm out}$ and
$\widetilde{\mathbf X}=\mathbf S_{\rm in}\mathbf X$, so that
\begin{equation}
\mathbf Y_{\rm cas}(\boldsymbol{\phi})
=
\mathbf W^T
\mathbf D_Q\mathbf S_Q
\mathbf D_{Q-1}\cdots
\mathbf S_2\mathbf D_1
\widetilde{\mathbf X}.
\label{eq:Y_cascade_interpretive}
\end{equation}
For layer $q$, define the partial cascades
\begin{equation}
\mathbf L_q
=
\mathbf W^T
\mathbf D_Q\mathbf S_Q
\cdots
\mathbf D_{q+1}\mathbf S_{q+1},
\qquad q<Q,
\end{equation}
with $\mathbf L_Q=\mathbf W^T$, and
\begin{equation}
\mathbf R_q
=
\mathbf S_q\mathbf D_{q-1}
\cdots
\mathbf S_2\mathbf D_1
\widetilde{\mathbf X},
\qquad q>1,
\end{equation}
with $\mathbf R_1=\widetilde{\mathbf X}$.
Writing
$\boldsymbol{\ell}_{q,k}=\mathbf L_q\mathbf e_k$
and
$\mathbf r_{q,k}^T=\mathbf e_k^T\mathbf R_q$,
the derivative of \eqref{eq:Y_cascade_interpretive} with respect to phase $\phi_{q,k}$ is
\begin{equation}
\frac{\partial\mathbf Y_{\rm cas}}{\partial\phi_{q,k}}
=
j e^{j\phi_{q,k}}
\boldsymbol{\ell}_{q,k}\mathbf r_{q,k}^T.
\label{eq:cascade_rank_one_direction}
\end{equation}
Here, $\mathbf r_{q,k}$ describes the incident field across the input
excitations at element $k$ of layer $q$, whereas
$\boldsymbol{\ell}_{q,k}$ describes its propagation toward the outputs.

The corresponding complex Jacobian atom is
\begin{equation}
\mathbf a_{q,k}
\triangleq
\operatorname{vec}
\left(
\frac{\partial\mathbf Y_{\rm cas}}{\partial\phi_{q,k}}
\right),
\label{eq:complex_atom_definition}
\end{equation}
which, using
$\operatorname{vec}(\mathbf a\mathbf b^T)=\mathbf b\otimes\mathbf a$,
becomes
\begin{equation}
\mathbf a_{q,k}
=
j e^{j\phi_{q,k}}
\left(
\mathbf r_{q,k}\otimes\boldsymbol{\ell}_{q,k}
\right).
\label{eq:cascade_complex_atom}
\end{equation}
Each phase thus generates a rank-one matrix perturbation, whose
vectorization is a Kronecker-structured sensitivity direction.

The real sensitivity vector is
\begin{equation}
\mathbf g_{q,k}
\triangleq
\begin{bmatrix}
\Re\{\mathbf a_{q,k}\}\\
\Im\{\mathbf a_{q,k}\}
\end{bmatrix}
\in\mathbb R^D.
\label{eq:real_atom_definition}
\end{equation}
Thus, the real Jacobian in \eqref{eq:real_unscaled_jacobian} can be written as
\begin{equation}
\mathbf J_R
=
\begin{bmatrix}
\mathbf J_1 & \cdots & \mathbf J_Q
\end{bmatrix},
\label{eq:empirical_frame_short}
\end{equation}
where
\begin{equation}
\mathbf J_q
=
\begin{bmatrix}
\mathbf g_{q,1} & \cdots & \mathbf g_{q,N}
\end{bmatrix}.
\end{equation}
The Jacobian frame therefore admits the layerwise decomposition
\begin{equation}
\mathbf J_R\mathbf J_R^T
=
\sum_{q=1}^{Q}\mathbf G_q,
\qquad
\mathbf G_q
=
\mathbf J_q\mathbf J_q^T
=
\sum_{k=1}^{N}
\mathbf g_{q,k}\mathbf g_{q,k}^T.
\end{equation}

Full row rank of $\mathbf J_R$  does not ensure that all transformation
directions are sufficiently sensitive to the phase controls.
To connect the layerwise decomposition to the singular-value analysis
of local synthesis developed above, consider
\begin{equation}
\begin{aligned}
\lambda_{\min}(\mathbf J_R\mathbf J_R^T)
&=
\min_{\|\mathbf u\|_2=1}
\sum_{q=1}^{Q}\mathbf u^T\mathbf G_q\mathbf u,\\
\lambda_{\max}(\mathbf J_R\mathbf J_R^T)
&=
\max_{\|\mathbf u\|_2=1}
\sum_{q=1}^{Q}\mathbf u^T\mathbf G_q\mathbf u.
\end{aligned}
\label{eq:layer_frame_extrema}
\end{equation}
The square roots of these eigenvalues are the weakest and strongest
aggregate sensitivities. Their ratio is small when at least one
transformation direction remains weak after combining all layers.
The layerwise decomposition shows that a direction poorly supported
by one layer can receive substantial contributions from others.
This possibility is general and does not depend on the cascade
approximation.

In the unilateral cascade, \eqref{eq:cascade_complex_atom} provides
a physical interpretation of this mechanism: propagation and phase
transformations before and after each layer reshape its sensitivity
directions. In the complete multiport model, the layer contributions
also account for electromagnetic coupling and multiple scattering,
but the same decomposition and spectral criteria remain applicable. Section~\ref{subsec:multiport_complementarity} tests this general
mechanism directly in the exact multiport Jacobian by measuring
whether the remaining layers compensate the weakest sensitivity
direction of each individual layer and whether this compensation
is accompanied by improved minimum sensitivity and conditioning.

%%%%%%%%%%%%%%%%%%%%%%%%%%%%%%%%%%%%%%%%%%%%%%%%%%%%%%%%%%%%%%%%%%%%%%%%%%%
\subsection{Geometric Limitations of a Single Layer and the Role of Depth}
\label{sec:single_layer_covariance}
%%%%%%%%%%%%%%%%%%%%%%%%%%%%%%%%%%%%%%%%%%%%%%%%%%%%%%%%%%%%%%%%%%%%%%%%%%%

For a single diagonal phase screen ($Q=1$, hence $P=N$),
\begin{equation}
\mathbf Y_{\rm cas}
=
\mathbf W^T\mathbf D\widetilde{\mathbf X}.
\end{equation}
Let $\widetilde{\mathbf x}_k^T$ denote row $k$ of
$\widetilde{\mathbf X}$ and $\mathbf w_k=\mathbf W^T\mathbf e_k$.
The complex Jacobian columns are
\begin{equation}
\mathbf a_k
=
j e^{j\phi_k}\mathbf c_k,
\qquad
\mathbf c_k
=
\widetilde{\mathbf x}_k\otimes\mathbf w_k.
\label{eq:single_atom_kron}
\end{equation}
The signatures $\mathbf c_k$ are fixed by the propagation geometry.
Writing $\mathbf J_c=[\mathbf a_1,\ldots,\mathbf a_N]$, their
complex Gramian is independent of the phase configuration:
\begin{equation}
\mathbf K
=
\mathbf J_c\mathbf J_c^H
=
\sum_{k=1}^{N}\mathbf c_k\mathbf c_k^H.
\label{eq:single_layer_phase_invariant_gramian}
\end{equation}
This invariance yields a geometric upper bound on the conditioning
of the real Jacobian
$\mathbf J_R=[(\Re\mathbf J_c)^T,(\Im\mathbf J_c)^T]^T$.

\medskip
\noindent\textit{Lemma (phase-independent conditioning bound).}
Let $\mathbf G(\boldsymbol{\phi})=\mathbf J_R\mathbf J_R^T$ and
assume $\mathbf K\neq\mathbf 0$. For every phase configuration,
\begin{equation}
\frac{\lambda_{\min}(\mathbf G(\boldsymbol{\phi}))}
     {\lambda_{\max}(\mathbf G(\boldsymbol{\phi}))}
\leq
\frac{\lambda_{\min}(\mathbf K)}
     {\lambda_{\max}(\mathbf K)}.
\label{eq:single_layer_conditioning_bound}
\end{equation}

\begin{IEEEproof}
For any unit vector $\mathbf u\in\mathbb C^{MI}$, define
$\mathbf u_R=[(\Re\mathbf u)^T,(\Im\mathbf u)^T]^T$.
The vectors $\mathbf u_R$ and $(j\mathbf u)_R$ are orthonormal and satisfy
\begin{equation}
\mathbf u_R^T\mathbf G\mathbf u_R
+
(j\mathbf u)_R^T\mathbf G(j\mathbf u)_R
=
\mathbf u^H\mathbf K\mathbf u.
\end{equation}
Choosing minimum- and maximum-eigenvalue eigenvectors of $\mathbf K$
and applying the Rayleigh quotient bounds gives
\begin{equation}
\lambda_{\min}(\mathbf G)
\leq \tfrac12\lambda_{\min}(\mathbf K),
\qquad
\lambda_{\max}(\mathbf G)
\geq \tfrac12\lambda_{\max}(\mathbf K),
\end{equation}
which proves \eqref{eq:single_layer_conditioning_bound}.
\end{IEEEproof}

The lemma is a direct consequence of spectral bounds properties.
Its architectural implication is that phase optimization cannot
overcome a small eigenvalue ratio of the geometry-dependent
matrix $\mathbf K$. Thus, a poorly conditioned spectrum of $\mathbf K$ represents a
structural geometric limitation: no phase configuration can make the
real Jacobian better conditioned than the bound in
\eqref{eq:single_layer_conditioning_bound}. 

The geometric bound vanishes for mirror-symmetric terminal
configurations. Consider identical, aligned terminal arrays
with $N_T$ elements each, whose corresponding transmitter and
receiver positions are mirror images across the SIM plane.
With individual transmitter excitations and direct receiver
observations, $I=M=N_T=N_R$. Under propagation preserving this
symmetry, $\mathbf W^T=\widetilde{\mathbf X}^{\,T}$, so
\begin{equation}
\mathbf c_k
=
\widetilde{\mathbf x}_k\otimes\widetilde{\mathbf x}_k
=
\operatorname{vec}\!\left(
\widetilde{\mathbf x}_k\widetilde{\mathbf x}_k^T
\right).
\end{equation}
Thus, all signatures lie in the complex symmetric subspace,
whose complex dimension is $N_T(N_T+1)/2$. Consequently,
\begin{equation}
\operatorname{rank}(\mathbf K)
\leq \min\{N,N_T(N_T+1)/2\}<N_T^2,
%\qquad N_T>1.
\end{equation}
for any $N_T>1$. Hence $\lambda_{\min}(\mathbf K)=0$, and
\eqref{eq:single_layer_conditioning_bound} implies
$\lambda_{\min}(\mathbf G(\boldsymbol{\phi}))=0$
for every phase configuration.

More precisely, phase multiplication preserves the symmetry
of each Jacobian atom, giving
\begin{equation}
\operatorname{rank}(\mathbf J_R)
\leq \min\{N,N_T(N_T+1)\}.
\label{eq:single_layer_symmetry_rank_bound}
\end{equation}
At least $N_T(N_T-1)$ real transformation directions are therefore
inaccessible, independently of the phases. 
%Consistently,
%\begin{equation}
%\mathbf Y_{\rm cas}
%=
%\widetilde{\mathbf X}^{\,T}\mathbf D\widetilde{\mathbf X}
%=
%\mathbf Y_{\rm cas}^{T}.
%\label{eq:single_layer_symmetric_channel}
%\end{equation}
%This restriction concerns the space of attainable transformations,
%not the matrix rank of $\mathbf Y_{\rm cas}$ itself.

For multiple layers, the incident and outgoing factors in
\eqref{eq:cascade_complex_atom} depend on the phases of the other
layers. Consequently, the complex Gramian generally depends on
the phase configuration, and the single-layer geometric bound
does not carry over as a fixed constraint.
In particular, symmetric terminal placement does not impose
the same restriction on a two-layer architecture.
For identical, aligned SIM grids with
$\mathbf S_2^T=\mathbf S_2$ and
$\mathbf W^T=\widetilde{\mathbf X}^{\,T}$,
\begin{equation}
\begin{aligned}
\mathbf Y_{\rm cas}
&=
\widetilde{\mathbf X}^{\,T}
\mathbf D_2\mathbf S_2\mathbf D_1
\widetilde{\mathbf X},\\
\mathbf Y_{\rm cas}^{T}
&=
\widetilde{\mathbf X}^{\,T}
\mathbf D_1\mathbf S_2\mathbf D_2
\widetilde{\mathbf X}.
\end{aligned}
\label{eq:two_layer_transpose}
\end{equation}
Unlike the single-layer expression, these two matrices need not coincide.
The same mechanism occurs for the Jacobian.
For example, at $\boldsymbol{\phi}=\mathbf 0$, one has 
\begin{equation}
\left.
\frac{\partial(\mathbf Y_{\rm cas}-\mathbf Y_{\rm cas}^{T})}
{\partial\phi_{2,k}}
\right|_{\boldsymbol{\phi}=\mathbf 0}
=
j\widetilde{\mathbf X}^{\,T}
(\mathbf E_k\mathbf S_2-\mathbf S_2\mathbf E_k)
\widetilde{\mathbf X}
\label{eq:two_layer_antisymmetric_derivative}
\end{equation}
where $\mathbf E_k=\mathbf e_k\mathbf e_k^T$.
Whenever \eqref{eq:two_layer_antisymmetric_derivative} is nonzero, a phase variation generates
sensitivity along an antisymmetric terminal direction that is
inaccessible in the symmetric single-layer configuration.
Thus, even a symmetric operating transformation (at $\boldsymbol{\phi}=\mathbf 0$ one has $\mathbf D_1=\mathbf D_2=\mathbf I$) can have Jacobian
directions extending beyond the symmetric subspace, when its layer
phases are independently controlled.

This mechanism requires inter-layer spatial mixing and suitable
terminal coupling. If $\mathbf S_2$ is diagonal, the phase screens
combine into a single diagonal operator and the single-layer symmetry
persists. More generally, the terminal projection in
\eqref{eq:two_layer_antisymmetric_derivative} may vanish for particular
geometries or controls.
Stacking therefore removes the universal symmetry constraint of the
single-layer configuration, but does not guarantee that all missing
directions are recovered or that the resulting Jacobian is well
conditioned.

The following example examines Jacobian conditioning as a function
of terminal distances, aperture, and layer count at comparable total
phase-control budgets.

%%%%%%%%%%%%%%%%%%%%%%%%%%%%%%%%%%%%%%%%%%%%%%%%%%%%%%%%%%%%%%%%%%%%%%%%%%%
\subsection{Numerical Illustration with Square SIM Layers}
\label{subsec:cascade_square_sim}

We illustrate the width--depth tradeoff using the scalar unilateral
cascade \eqref{eq:Y_cascade_interpretive}, with discrete square
SIM grids on parallel planes in three-dimensional space.
The transmitter and receiver each consist of four elements arranged on
a centered $2\times2$ grid. Individual transmitter excitations and
direct receiver outputs are considered, so that $I=M=4$.
Their transverse coordinates are all combinations
$(\pm D_T/2,\pm D_T/2)$ and $(\pm D_R/2,\pm D_R/2)$, respectively,
with $D_T=D_R=\lambda/2$.

Taking the first SIM plane as longitudinal coordinate zero, the Tx
plane is at $-d_T$, layer $q$ is at $(q-1)d_{\rm layer}$, and the Rx
plane is at $(Q-1)d_{\rm layer}+d_R$, where
$d_{\rm layer}=\lambda$ and $Q\in\{1,2,3,4\}$.

All layers in a given configuration have identical, aligned square
grids with $K_s$ elements per side and $N=K_s^2$ independently
controlled phases per layer. The transverse element coordinates are
\begin{equation}
\begin{gathered}
\{(s_i,s_j):i,j=0,\ldots,K_s-1\},\\
s_i=\left(i-\frac{K_s-1}{2}\right)\Delta,~ s_j=\left(j-\frac{K_s-1}{2}\right)\Delta,
\end{gathered}
\end{equation}
where $K_s$ is odd and $\Delta=\lambda/2$.
The aperture parameter $A=(K_s-1)\Delta$ denotes the side length
between the outermost element centers.

External propagation, from the transmitter array to the first SIM
layer and from the last SIM layer to the receiver array, is described
by a scalar spherical-wave model. The coupling coefficient between
two terminal and SIM elements separated by a distance $r$ is
\begin{equation}
h_{\rm ext}(r)=\frac{e^{-jk_0r}}{r},
\qquad k_0=\frac{2\pi}{\lambda},
\label{eq:spherical_channels}
\end{equation}
up to a common multiplicative constant. This model retains the
distance-dependent phase and spherical amplitude spreading, using
the actual Euclidean distance for each pair of elements.

Propagation between adjacent SIM layers follows the
Rayleigh--Sommerfeld diffraction model \cite{10158690}:
\begin{equation}
[\mathbf S_q]_{ij}
=
\Delta^2
\frac{\cos\theta_{ij}}{r_{ij}}
\left(
\frac{1}{2\pi r_{ij}}+\frac{j}{\lambda}
\right)
e^{-jk_0r_{ij}},
\label{eq:RS_interlayer}
\end{equation}
where $r_{ij}$ is the distance between the two elements,
$\theta_{ij}$ is the angle between their connecting direction
and the normal to the layers, and $\Delta^2$ is the spatial
sampling area. 

The single-layer reference has side length $A_{\rm ref}=25\lambda$
and half-wavelength element spacing, giving $K_{s,\rm ref}=51$
elements per side and $P_{\rm ref}=51^2=2601$ phase controls.
For each layer count, $K_s$ is obtained by rounding
$\sqrt{P_{\rm ref}/Q}$ to the nearest integer and increasing it
by one if the result is even.
For $Q=1,2,3,4$, this gives $K_s=51,37,29,27$, respectively,
corresponding to side lengths $A/\lambda=25,18,14,13$ and total
phase-control budgets $P=QK_s^2= 2601,2738,2523,2916$.
The total phase-control budget is kept approximately constant,
up to rounding required by the discrete grid sizes.

Let $\mathbf J_R\in\mathbb R^{32\times P}$ be the realified
transformation Jacobian in
\eqref{eq:empirical_frame_short}.
The reported conditioning metric is $\gamma$ in \eqref{eq:GN_relative_conditioning}.
%\begin{equation}
%\gamma
%=
%\frac{\sigma_{\min}(\mathbf J)}
%{\sigma_{\max}(\mathbf J)}.
%\label{eq:cascade_square_conditioning}
%\end{equation}
All 32 singular values are included, so rank deficiency gives
$\gamma=0$, whereas larger values indicate more balanced sensitivities.
For each geometry, 20 independent phase profiles are drawn with all
element phases uniformly distributed in $[0,2\pi)$, and the resulting values of
$\gamma$ are averaged.

\begin{figure}[!tp]
\centering
\includegraphics[width=0.95\columnwidth]{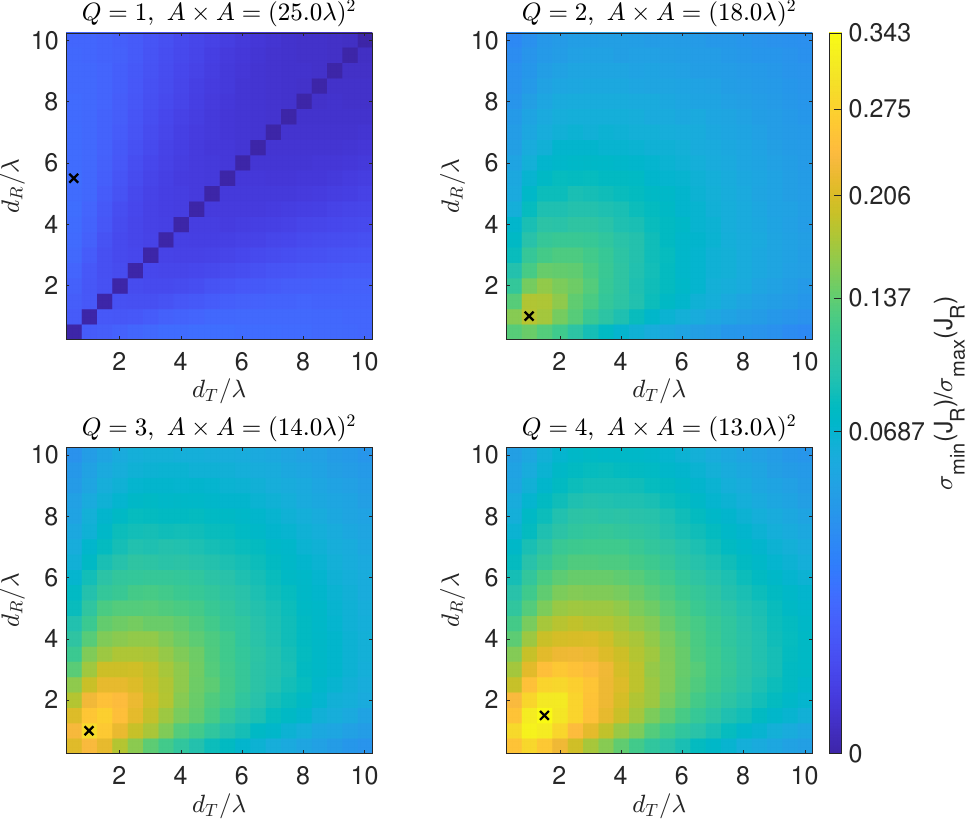}
\caption{Mean reciprocal Jacobian conditioning $\gamma$ for square
SIMs with $Q=1,2,3,4$ at approximately equal total control budget.
The indicated $A$ is the side length of each layer.
Each pixel averages 20 random phase profiles; crosses mark the
maximum values.
All panels use the same color mapping for the values of $\gamma$.}
\label{fig:cascade_heatmaps_2D}
\end{figure}

Figure~\ref{fig:cascade_heatmaps_2D} shows the mean conditioning as a function of 
$d_T/\lambda$ and $d_R/\lambda$ in the range $[0.5,10]$, for the 4 considered architectures.
The largest value of $\gamma$ for each architecture is marked with a cross.
It is apparent that conditioning steadily improves with the number of layers $Q$. This occurs uniformly with respect to the spatial dimensions $d_T$, $d_R$. In particular, the single-layer structure shows a uniformly poor conditioning, while a significant improvement is observed for $Q=2$. This confirms the intuition suggested by the discussion in Section \ref{sec:cascade_jacobian_interpretation}.

\begin{figure}[!tp]
\centering
\includegraphics[width=0.95\columnwidth]{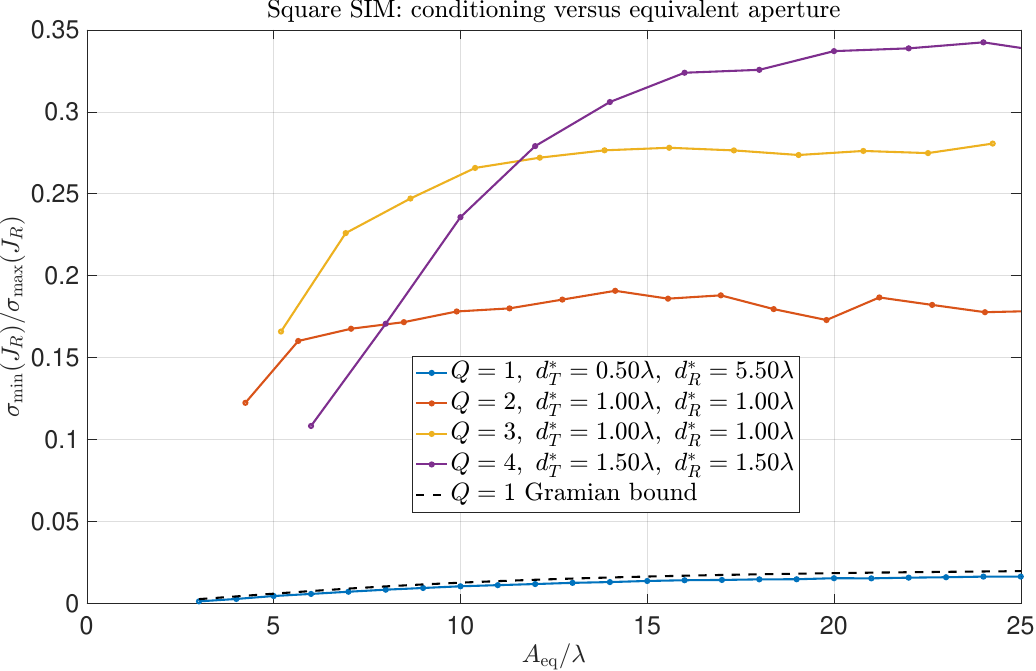}
\caption{Mean reciprocal Jacobian conditioning versus equivalent
aperture $A_{\rm eq}/\lambda$, on linear axes.
Each point averages 20 random phase profiles.
For each $Q$, the terminal distances shown in the legend are selected
from the maximum-budget heatmap and held fixed throughout the sweep.
For $Q=1$, the phase-independent Gramian bound derived in
\eqref{eq:single_layer_conditioning_bound} is also shown.}
\label{fig:cascade_aperture_2D}
\end{figure}

The single-layer heatmap displays a zero-valued diagonal at $d_T=d_R$,
as predicted by the symmetry argument in Section \ref{sec:single_layer_covariance}.
In this configuration, $\mathbf Y_{\rm cas}$ is 
a complex $4 \times 4$ symmetric matrix containing at most $10$ independent complex entries,
or $20$ real components, whereas the complete transformation space
has $32$ real dimensions. Consequently, $\operatorname{rank}(\mathbf J_R)\leq20$ and
$\gamma=0$.
Thus, the heatmap diagonal reflects a structural loss of sensitivity, not an
insufficient phase-control budget.
Away from $d_T=d_R$, symmetry is lost 
and the single-layer heatmap exhibits nonzero conditioning values.
The multilayer panels also exhibit nonzero values along the
equal-distance diagonal, consistently with the additional
antisymmetric sensitivities permitted by inter-layer propagation
and independent phase control.

{In order to evaluate the effect of the SIM aperture $A$, 
the conditioning analysis has been repeated by considering every odd value of $K_s$ from $7$ to the
value considered so far for the corresponding $Q$.
For each $Q$, the optimal values of  $d_T$, $d_R$, corresponding to the crosses in Figure ~\ref{fig:cascade_heatmaps_2D}, are considered. For each $(Q,A)$ pair, the average conditioning is evaluated on 20 random phase profiles. The results are shown in Figure~\ref{fig:cascade_aperture_2D}.
Mean conditioning is reported as a function of the equivalent aperture $A_{eq}/\lambda$, with $A_{\rm eq}=\sqrt{Q}\,A$, so to have a common scale across architectures with different number of layers.
The single-layer curve flattens near $0.02$. Figure~\ref{fig:cascade_aperture_2D}
also reports the phase-independent Gramian bound derived in
\eqref{eq:single_layer_conditioning_bound}. Remarkably, this bound
closely follows the actual mean conditioning over the entire aperture
range, showing that the poor conditioning observed for $Q=1$ is
essentially dictated by the propagation geometry rather than by an
unfavorable choice of the phase profile.
}

The experiments expose the structural sensitivity loss predicted for
a mirror-symmetric single layer: increasing its aperture cannot recover
the excluded antisymmetric transformation directions. Multiple layers
remove this constraint and, in the tested geometries, provide
substantially better conditioning that improves with depth and can
benefit from larger apertures. 

%%%%%%%%%%%%%%%%%%%%%%%%%%%%%%%%%%%%%%%%%%%%%%%%%%%%%%%%%%%%%%%%%%%%%%%%%%%
\section{Multiport Transformation Jacobian}
\label{sec:multiport_jacobian}
%%%%%%%%%%%%%%%%%%%%%%%%%%%%%%%%%%%%%%%%%%%%%%%%%%%%%%%%%%%%%%%%%%%%%%%%%%%

The cascaded representation has been used above to expose the structure and conditioning of the transformation directions generated by the programmable phases. We now evaluate the corresponding local sensitivities directly from the complete S-parameter multiport model of Section~\ref{sec:S_multiport_model}. This allows the same expressivity indicators to be assessed while accounting for reciprocal coupling, multiple reflections, and the load-dependent redistribution of the electromagnetic field.

Starting from \eqref{eq:Y_compact_S},
\begin{equation}
\mathbf Y(\boldsymbol{\phi})=\mathbf Y_d+\mathbf A\boldsymbol{\Gamma}(\boldsymbol{\phi})\mathbf F(\boldsymbol{\phi})\mathbf B,
\end{equation}
with
\begin{equation}
\mathbf F(\boldsymbol{\phi})=\left(\mathbf I-\mathbf S_{EE}\boldsymbol{\Gamma}(\boldsymbol{\phi})\right)^{-1},
\end{equation}
define
\begin{equation}
\boldsymbol{\Gamma}_p'=\frac{\partial\boldsymbol{\Gamma}(\boldsymbol{\phi})}{\partial\phi_p}.
\end{equation}
Differentiation of the inverse gives
\begin{equation}
\frac{\partial\mathbf F}{\partial\phi_p}=\mathbf F\mathbf S_{EE}\boldsymbol{\Gamma}_p'\mathbf F.
\end{equation}
Hence,
\begin{equation}
\frac{\partial\mathbf Y}{\partial\phi_p}
=
\mathbf A
\left(
\mathbf I+\boldsymbol{\Gamma}\mathbf F\mathbf S_{EE}
\right)
\boldsymbol{\Gamma}_p'
\mathbf F\mathbf B.
\end{equation}
Using
\begin{equation}
\mathbf I+\boldsymbol{\Gamma}\mathbf F\mathbf S_{EE}=
\left(\mathbf I-\boldsymbol{\Gamma}\mathbf S_{EE}\right)^{-1},
\end{equation}
the exact sensitivity can be written compactly as
\begin{equation}
\frac{\partial\mathbf Y}{\partial\phi_p}
=
\mathbf A
\left(\mathbf I-\boldsymbol{\Gamma}\mathbf S_{EE}\right)^{-1}
\boldsymbol{\Gamma}_p'
\mathbf F\mathbf B.
\label{eq:dY_multiport}
\end{equation}

Equation~\eqref{eq:dY_multiport} retains the left--local perturbation--right factorization highlighted by the ideal cascade. The difference is that the two surrounding factors now contain the solution of the complete coupled electromagnetic network. A perturbation of one programmable cell can therefore modify the transformation through reciprocal coupling and repeated scattering rather than only through the forward cascade paths. For the block-diagonal cell structure described in Section~\ref{sec:S_multiport_model}, $\boldsymbol{\Gamma}_p'$ is nonzero only within the block associated with the $p$th programmable cell. This structure can be exploited to compute the sensitivities efficiently, as in the low-complexity optimization gradient descent approach developed in \cite{Abra_SIM1,PettaniceSIM2026}.

As in the cascaded analysis, vectorizing \eqref{eq:dY_multiport}  gives
\begin{equation}
\mathbf J_c=
\begin{bmatrix}
\displaystyle\frac{\partial\operatorname{vec}(\mathbf Y)}{\partial\phi_1}&\cdots&
\displaystyle\frac{\partial\operatorname{vec}(\mathbf Y)}{\partial\phi_P}
\end{bmatrix},
\end{equation}
and the corresponding real Jacobian is
\begin{equation}
\mathbf J_R=
\begin{bmatrix}
\Re\{\mathbf J_c\}\\
\Im\{\mathbf J_c\}
\end{bmatrix}
\in\mathbb R^{D\times P}.
\label{eq:JR}
\end{equation}
%For the local inverse-synthesis problem of Section~\ref{sec:optimization}, the optimal coefficient $\beta_\star$ is held fixed while constructing the phase update. Hence, the Jacobian appearing in \eqref{eq:real_synthesis_jacobian} is related to the physical transformation Jacobian by
%\begin{equation}
%\mathbf J
%=
%\begin{bmatrix}
%\Re\{\beta_\star\mathbf J_c\}\\
%\Im\{\beta_\star\mathbf J_c\}
%\end{bmatrix}.
%\label{eq:J_from_Jc}
%\end{equation}
%Multiplication by the nonzero complex scalar $\beta_\star$ corresponds, after realification, to a scaled orthogonal transformation. Therefore,
%\begin{equation}
%\sigma_i(\mathbf J)=|\beta_\star|\,\sigma_i(\mathbf J_R),
%\label{eq:J_JR_singular_values}
%\end{equation}
%so that $\mathbf J$ and $\mathbf J_R$ have the same rank, condition number, and normalized singular-value spectrum. The architectural conditioning analysis can therefore be carried out directly on $\mathbf J_R$, whereas $\mathbf J$ is the corresponding Jacobian entering the local inverse-synthesis update.
%
%{
%The previous discussion on the Jacobian conditioning introduced for the cascaded model, applies directly also to the multiport Jacobian \eqref{eq:JR}: the only difference is that the physical transformation directions used to construct its columns are now obtained from the complete electromagnetic model.
%}

%%%%%%%%%%%%%%%%%%%%%%%%%%%%%%%%%%%%%%%%%%%%%%%%%%%%%%%%%%%%%%%%%%%%%%%%%%%
\section{Numerical Results}
\label{sec:results}
%%%%%%%%%%%%%%%%%%%%%%%%%%%%%%%%%%%%%%%%%%%%%%%%%%%%%%%%%%%%%%%%%%%%%%%%%%%

\subsection{Simulation Setup and Performance Metrics}
\label{sec:simulation_setup}

We now numerically assess the expressivity properties discussed in the previous sections by comparing SIM architectures having substantially different width and depth while preserving the same overall number of programmable phase variables. All the results are obtained using the complete multiport electromagnetic model introduced in Section~\ref{sec:S_multiport_model}, with the corresponding Jacobian derived in Section \ref{sec:multiport_jacobian}.

The physical implementation follows the dipole-based model of \cite{Abra_SIM1}. The carrier frequency is $f_0=28$~GHz, and each electromagnetic port is represented by a thin metallic dipole oriented along the $z$ direction, with length $L=0.46\lambda$ and radius $\lambda/500$. The dipoles of each layer form a uniform planar array in the $yz$ plane, with inter-element spacings
$d_y=\frac{\lambda}{2}$, $d_z=\frac{3\lambda}{4}$,
whereas the $x$ axis identifies the longitudinal direction across the SIM. Consecutive physical layers are separated by $d_x=\lambda$.

The electromagnetic interaction between the dipoles belonging to each pair of coupled adjacent layers is evaluated through the self and mutual impedances using the analytical procedure of \cite{DR1}, as in \cite{Abra_SIM1}. The diagonal entries of the resulting impedance matrices account for the self impedances of the individual dipoles, whereas the off-diagonal entries describe their mutual electromagnetic coupling. The corresponding scattering matrices are obtained through the standard impedance-to-scattering transformation
$
\mathbf S=
\left(\mathbf Z-Z_0\mathbf I\right)
\left(\mathbf Z+Z_0\mathbf I\right)^{-1},
$
where $Z_0$ denotes the reference port impedance. The resulting layered multiport representation therefore retains intra-layer coupling, reciprocal forward and backward coupling between consecutive layers, and the associated multiple-scattering effects. Each physical layer comprises $N=L_yL_z$ programmable cells, with $L_y$ and $L_z$ denoting the numbers of cells along the $y$ and $z$ directions, respectively. All architectures are compared under the same overall phase budget
\begin{equation}
P=Q L_yL_z=6144,
\end{equation}
where $Q$ denotes the number of programmable layers. Hence, increasing depth necessarily reduces the number of programmable cells per layer, allowing the width--depth tradeoff to be investigated without changing the total number of controllable real parameters.

For compactness, each architecture is denoted as
\begin{equation}
\mathsf{W}[L_y{\times}L_z]
\qquad\text{or}\qquad
\mathsf{D}[L_y{\times}L_z,Q],
\end{equation}
where $\mathsf{W}$ denotes a wide single-layer architecture, for which $Q=1$ is implicit, whereas $\mathsf{D}$ denotes a multilayer deep architecture and therefore explicitly includes the number of layers $Q$. The wide configurations considered are
\[
\mathsf{W}[768{\times}8],\quad
\mathsf{W}[384{\times}16],\quad
\mathsf{W}[192{\times}32],\quad
\mathsf{W}[96{\times}64],
\]
while the corresponding deep configurations are
\[
\mathsf{D}[96{\times}8,8],\quad
\mathsf{D}[64{\times}8,12],\quad
\mathsf{D}[64{\times}16,6],\quad
\mathsf{D}[32{\times}16,12].
\]
All these architectures satisfy $P=6144$ and therefore differ only in how the available programmable degrees of freedom are distributed between the transverse dimensions and the depth of the SIM.

Two classes of input excitations are considered. In the far-field (FF) scenario, the columns of the input matrix $\mathbf X$ are spatial DFT modes, following the angular-processing configuration considered in \cite{Abra_SIM1}. We consider $N_\theta=16$ and $N_\theta=32$ spatial modes along the $y$ direction and one mode along $z$, yielding the scenarios denoted as $\mathsf{FF}(16)$ and $\mathsf{FF}(32)$. The corresponding illumination vectors are uniformly sampled in spatial frequency over the DFT grid supported by the first SIM layer, thereby probing distinct far-field angular directions.

In the near-field (NF) scenario, we instead consider the localization geometry investigated in \cite{PettaniceSIM2026,abrardo_Eusipco}. The transmitters are placed on an angle--range grid comprising $N_\theta$ angular positions and $N_r$ radial distances, yielding
$
I=N_\theta N_r
$
input configurations. The two scenarios considered are denoted as $\mathsf{NF}(8,2)$ and $\mathsf{NF}(8,4)$, corresponding to $(N_\theta,N_r)=(8,2)$ and $(8,4)$, respectively. In this case, the columns of $\mathbf X$ are generated by the spherical-wave fields produced by the corresponding transmitter positions and therefore jointly encode angular and range information.

The output ports are implemented by receiving probes having the same physical characteristics as the dipoles constituting the SIM, following the setup adopted in \cite{Abra_SIM1}. Specifically, the output probes are $z$-oriented dipoles with the same length and radius as the SIM elements and are placed on a line parallel to the $y$ axis at a distance $\lambda$ from the last SIM layer. Adjacent output probes are separated by $\lambda/2$ along the $y$ direction.

For all scenarios, $M=I$, giving real output dimension $D=2MI$. These Tx/Rx arrangements provide the standard angular- and position-discrimination sensing reference considered in \cite{Abra_SIM1,PettaniceSIM2026,abrardo_Eusipco}. They specify the physical input and output geometry; the Jacobian experiments below do not require a prescribed target transformation.

Besides the synthesis error, the local expressivity of each architecture is characterized through the normalized singular-value spectrum of the real transformation Jacobian $\mathbf J_R$. For a prescribed relative threshold $\tau$, we define the effective rank
\begin{equation}
r_{\rm eff}(\tau)
=
\#\left\{
i:
\frac{\sigma_i(\mathbf J_R)}
{\sigma_{\max}(\mathbf J_R)}
>\tau
\right\},
\label{eq:reff_results}
\end{equation}
and the corresponding normalized effective rank
\begin{equation}
\rho_J(\tau)
=
\frac{r_{\rm eff}(\tau)}{D}.
\label{eq:normalized_rank_results}
\end{equation}
Hence, $\rho_J(\tau)=1$ indicates that all the Jacobian directions remain above the considered relative threshold. For example,  $\rho_J(10^{-5})=1$ means that all the $D$ singular values lie within five orders of magnitude of the largest one. Conversely, $\rho_J(10^{-5})<1$ indicates that a fraction of these Jacobian directions falls below this threshold. In this sense, $\rho_J(\tau)$ provides a compact numerical description of the singular-value distribution, complementing both the Jacobian rank and its condition number.

\subsection{Random-Phase Multiport Jacobian}
\label{subsec:numerical_figures}

\begin{figure}[!tp]
\centering
\includegraphics[width=\columnwidth]{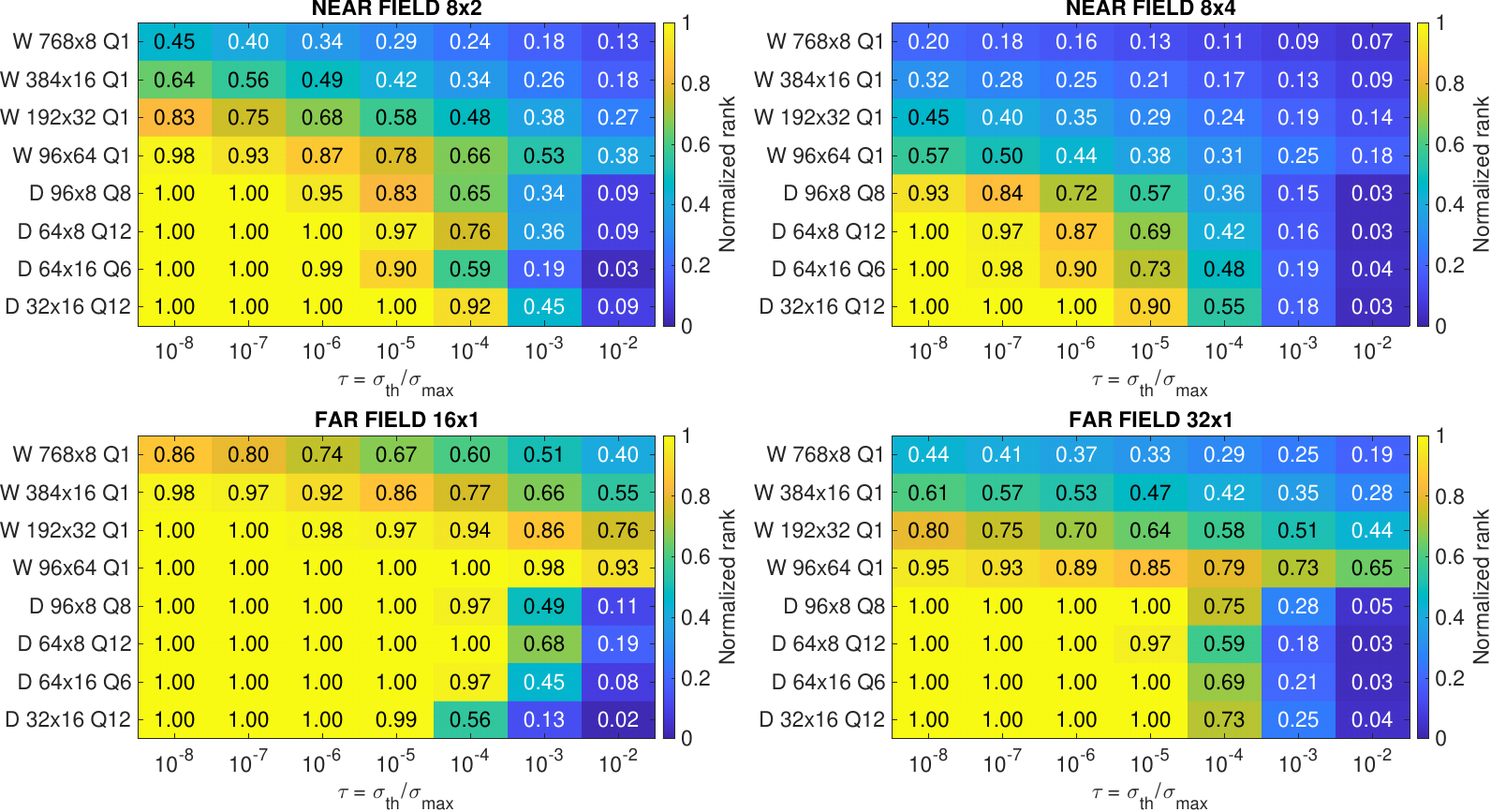}
\caption{Mean normalized effective rank over $100$ independent random phase configurations, using the exact multiport Jacobian without optimization. Panels: (a) $\mathsf{NF}(8,2)$, (b) $\mathsf{NF}(8,4)$, (c) $\mathsf{FF}(16)$, and (d) $\mathsf{FF}(32)$. Colors and cell values report the sample mean of $\rho_J(\tau)$.}
\label{fig:rank_4cases}
\end{figure}

We first evaluate the exact multiport Jacobian at $100$ independent random phase configurations for each architecture and scenario, without performing any optimization. Figure~\ref{fig:rank_4cases} reports the sample mean of $\rho_J(\tau)$. Its four panels correspond to $\mathsf{NF}(8,2)$, $\mathsf{NF}(8,4)$, $\mathsf{FF}(16)$, and $\mathsf{FF}(32)$, respectively. The horizontal axis gives the relative singular-value threshold and each row identifies an architecture. This experiment isolates the local sensitivity spectrum from both the target and the optimization trajectory.

Distributing the fixed phase budget across multiple layers systematically improves the coverage of weak output directions in the demanding NF and $\mathsf{FF}(32)$ scenarios: the deep architectures retain a larger fraction of directions at the smaller relative thresholds. The improvement is particularly evident for $\mathsf{D}[32{\times}16,12]$ in the NF cases. This is not a uniform ordering at every threshold or a monotonic dependence on layer count alone. At larger thresholds, some wide architectures retain a larger fraction of strong directions, and $\mathsf{W}[96{\times}64]$ already exhibits a favorable spectrum in $\mathsf{FF}(16)$. Thus the maps support an architectural benefit in recovering weak sensitivities, while also showing the role of aperture shape and propagation geometry. 

\subsection{Synthesis of a Reachable Random Target}

To distinguish numerical difficulty from target infeasibility, for each architecture we generate $\mathbf T=\mathbf Y(\boldsymbol{\phi}_{\rm target})$ from a random phase configuration of the same physical model and initialize synthesis at an independent random configuration. Each target is therefore reachable by construction; the comparison concerns recovery of architecture-specific reachable targets, rather than synthesis of one common transformation. The normalized synthesis error in~\eqref{eq:SI_cost} is minimized using Gauss--Newton with Armijo backtracking. At each iteration, the phase update is
\begin{equation}
\boldsymbol{\phi}_{k+1}
=
\boldsymbol{\phi}_k
+
\alpha_k\Delta\boldsymbol{\phi}_{\mathrm{GN},k},
\end{equation}
where the step factor $\alpha_k\in(0,1]$ is reduced by backtracking until the proposed phase update satisfies the Armijo sufficient-decrease condition.

\begin{figure}[!tp]
\centering
\includegraphics[width=0.95\columnwidth]{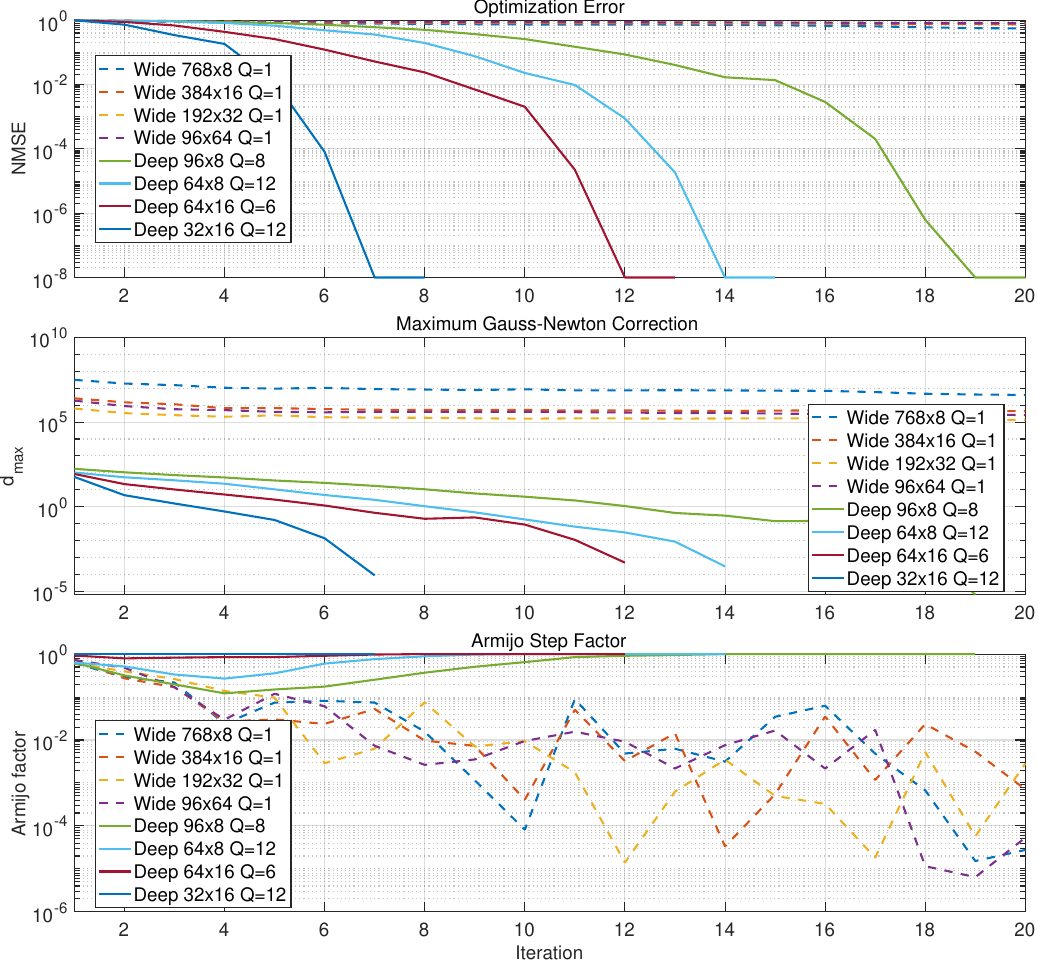}
\caption{GN synthesis in the $\mathsf{NF}(8,4)$ scenario of targets generated by random configurations of the corresponding architecture, starting from independent random configurations. Curves are averages over $20$ random trials per architecture. From top to bottom: synthesis NMSE, $d_{\max}$ in~\eqref{eq:dmax_results}, and Armijo factor versus iteration.}
\label{fig:reachable_target}
\end{figure}

We then consider the $\mathsf{NF}(8,4)$ scenario, the most challenging among the four geometries examined. Figure~\ref{fig:reachable_target} summarizes $20$ random trials per architecture, each with a reachable target and an independent random initialization; the curves report averages over these trials. The first panel shows the synthesis NMSE versus iteration. Only the deep configurations converge to the displayed accuracy floor, whereas the wide configurations exhibit little error reduction over the displayed iteration budget. Since the target is reachable by construction in every trial for every architecture, this contrast cannot be attributed to global infeasibility: it highlights the numerical difficulty of recovering a realizable transformation with the wide configurations.

The second panel shows the largest inverse singular coefficient of the raw Gauss-Newton correction. Using the notation  introduced in Section~\ref{sec:optimization}, at each iteration of each trial we define
\begin{equation}
d_{\max}=\max_{i:\,\sigma_i>0}\left|\frac{c_i}{\sigma_i}\right|.
\label{eq:dmax_results}
\end{equation}
where $c_i=\mathbf u_i^T\mathbf e$ are the residual projections and $\sigma_i$ the corresponding  singular values of Jacobian $\mathbf J_R$.
The plotted diagnostic is the average of these per-trial maxima.
According to \eqref{eq:GN_correction_norm}, it measures the largest correction coefficient in the right-singular-vector basis and accounts for both weak sensitivity and the residual component requiring correction along that direction. 
The deep configurations start with much smaller $d_{\max}$ and reduce it as synthesis progresses, whereas the wide configurations retain very large values. Thus, their weak singular directions carry residual components whose correction requires large phase-space displacements, potentially outside the local linear regime.

The third panel shows the Armijo factor versus iteration, averaged over the same trials. The factors remain close to unity for the deep configurations, indicating that backtracking imposes little additional reduction of the proposed steps, whereas substantial reductions are required for the wide configurations. Together, the second and third panels connect the observed NMSE behavior to the local inverse problem: deep architectures support moderate corrections that are largely accepted, while wide architectures require large corrections along weak directions and accept strongly reduced steps, hindering progress despite target reachability.

\subsection{Multiport Test of Layer Complementarity}
\label{subsec:multiport_complementarity}

We next test the compensation mechanism directly in the exact real multiport Jacobian. Partition its columns by layer as $\mathbf J_R=[\mathbf J_1\ \cdots\ \mathbf J_Q]$, and let $\mathbf J_{-q}$ concatenate all blocks except $\mathbf J_q$. Then
\begin{equation}
\mathbf J_R\mathbf J_R^T
=\mathbf J_q\mathbf J_q^T+\mathbf J_{-q}\mathbf J_{-q}^T.
\end{equation}
For a full-row-rank block $\mathbf J_q$, let $\mathbf u_{q,\min}$ be a unit left singular vector associated with its minimum singular value. Define
\begin{equation}
\xi_q^2=
\frac{\mathbf u_{q,\min}^T\mathbf J_{-q}\mathbf J_{-q}^T\mathbf u_{q,\min}}
{\sigma_{\min}^2(\mathbf J_q)}.
\label{eq:multiport_gminus}
\end{equation}
Thus $\xi_q=\|\mathbf J_{-q}^T\mathbf u_{q,\min}\|_2/\sigma_{\min}(\mathbf J_q)$ measures the sensitivity of the other layers along the weakest direction of layer $q$. 
Assuming that both $\mathbf J_q$ and $\mathbf J_{-q}$ have full
row rank, define the minimum- and maximum-sensitivity gains as
\begin{equation}
\Gamma_q
=\frac{\sigma_{\min}(\mathbf J_{-q})}
{\sigma_{\min}(\mathbf J_q)},
\qquad
\Lambda_q
=\frac{\sigma_{\max}(\mathbf J_{-q})}
{\sigma_{\max}(\mathbf J_q)}.
\label{eq:multiport_gamma_lambda_minus}
\end{equation}
%Their ratio then satisfies
%\begin{equation}
%C_q^{(-q)}
%=\frac{\Gamma_q}{\Lambda_q}
%=\frac{\operatorname{cond}(\mathbf J_q)}
%{\operatorname{cond}(\mathbf J_{-q})}.
%\label{eq:multiport_Cminus}
%\end{equation}
Hence $\Gamma_q>\Lambda_q$ means that the minimum sensitivity increases proportionally more than the maximum and that $\mathbf J_{-q}$ is better conditioned than $\mathbf J_q$. 

\begin{figure}[!tp]
\centering
\includegraphics[width=\columnwidth]{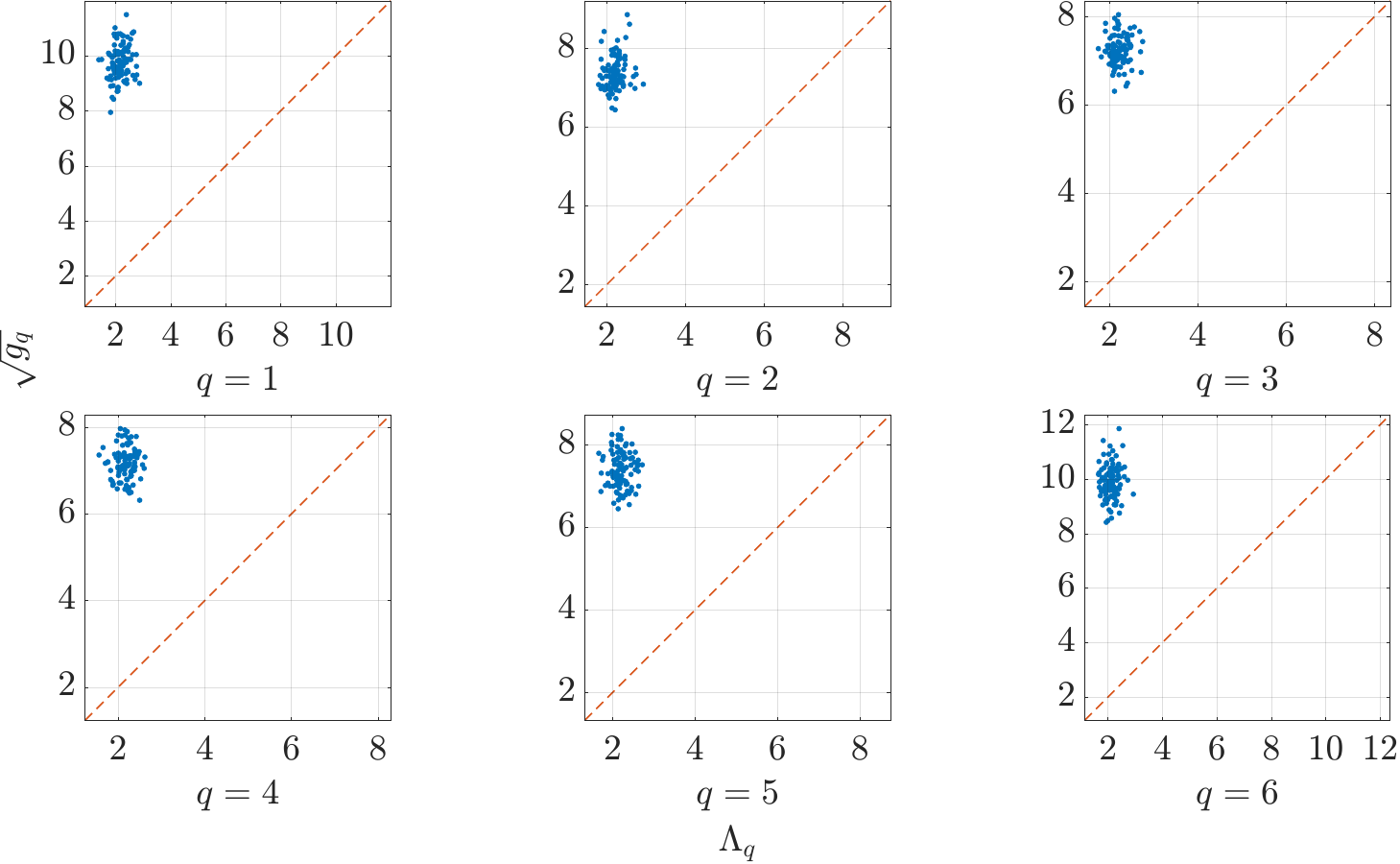}
\caption{Directional complementarity: $\xi_q$ versus $\Lambda_q$. Each panel corresponds to one layer of $\mathsf{D}[64{\times}16,6]$ in $\mathsf{NF}(8,2)$ and contains $100$ random configurations. The dashed line is $y=x$.}
\label{fig:complementarity_directional}
\end{figure}

\begin{figure}[!tp]
\centering
\includegraphics[width=\columnwidth]{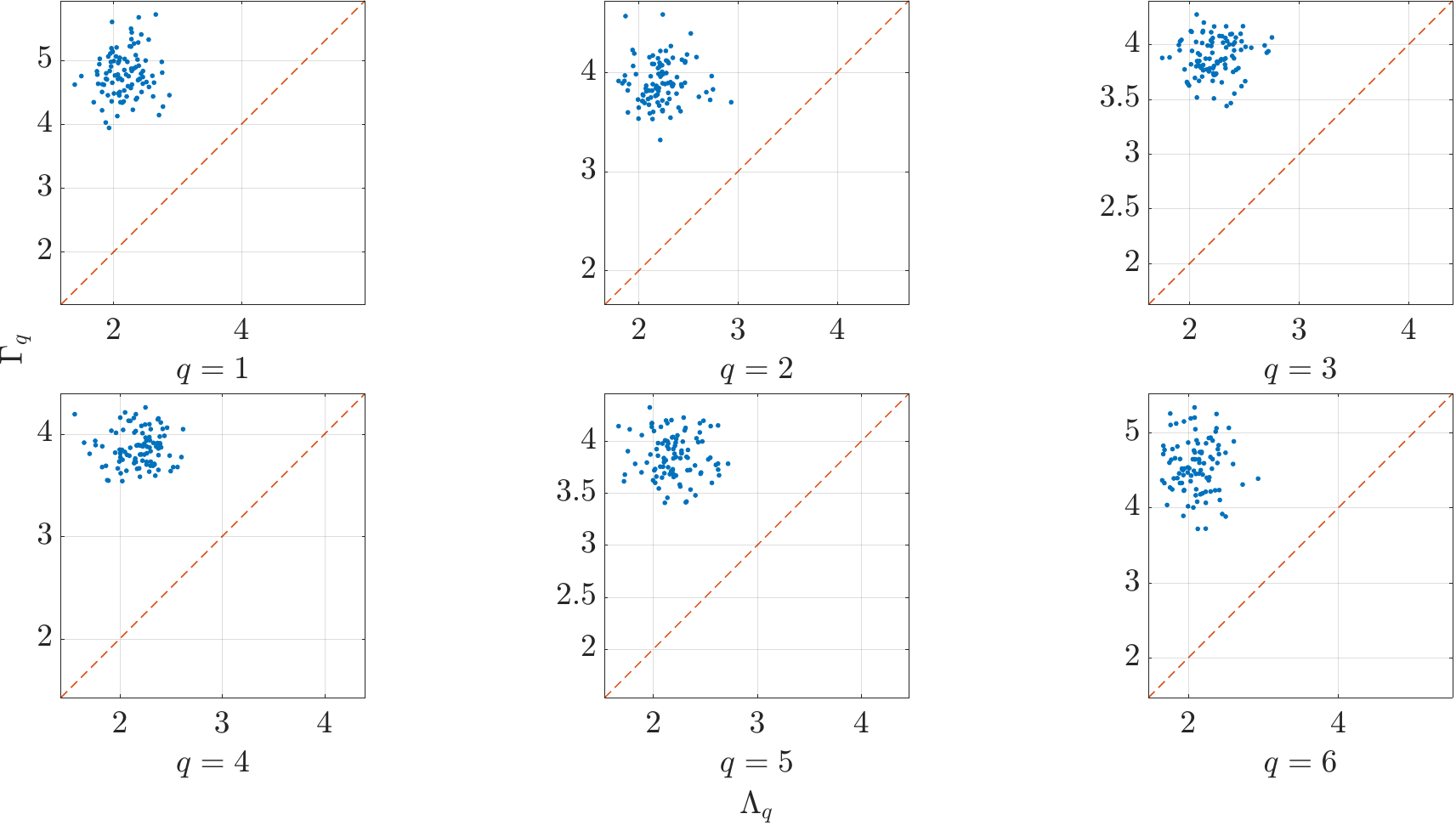}
\caption{Minimum- versus maximum-sensitivity gains: $\Gamma_q$ versus $\Lambda_q$, for the same layer-resolved samples. Points above $y=x$ indicate improved conditioning of $\mathbf J_{-q}$ relative to $\mathbf J_q$.}
\label{fig:complementarity_conditioning}
\end{figure}

The experiment uses $\mathsf{D}[64{\times}16,6]$ in $\mathsf{NF}(8,2)$, with $D=512$ and $N=1024$ tuning variables per layer, so that each $\mathbf J_q\in\mathbb R^{512\times1024}$ can have full row rank. The same $100$ random configurations used in the random-phase analysis are evaluated without optimization. Figures~\ref{fig:complementarity_directional}--\ref{fig:complementarity_conditioning} each contain six layer-resolved panels for $q=1,\dots,6$, with $100$ paired observations per panel.

Figure~\ref{fig:complementarity_directional} shows that $\xi_q$ exceeds $\Lambda_q$ for all six layers. This means that the remaining layers compensate the weak direction of layer $q$, beyond an overall increase in Jacobian magnitude. Figure~\ref{fig:complementarity_conditioning} shows that  $\Gamma_q$ consistently exceeds $\Lambda_q$, confirming that this compensation accompanies a larger gain in minimum than maximum sensitivity and consequently improved conditioning.
These results support layer complementarity in the complete multiport model.

\subsection{Identity-Target Synthesis as an Additional Check}

Finally, we consider the sensing-related transformation
\begin{equation}
\mathbf T=\mathbf I_I.
\label{eq:identity_target_results}
\end{equation}
It associates each input hypothesis with one output port while suppressing the others, providing a canonical identification/separation task: angular discrimination in FF and position discrimination in NF. Unlike the preceding random targets, this transformation is not guaranteed reachable by construction.

\begin{figure}[!tp]
\centering
\includegraphics[width=0.48\textwidth]{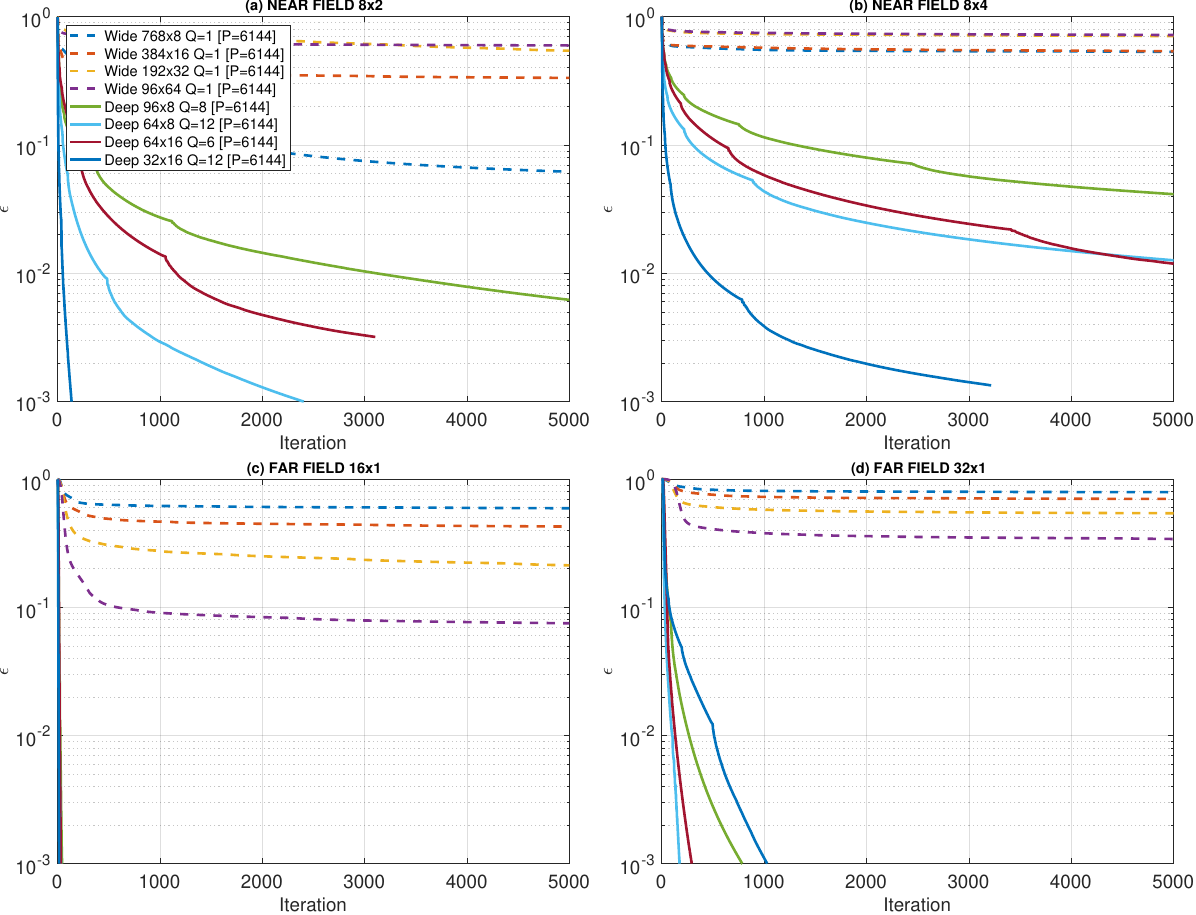}
\caption{Identity-target gradient-descent convergence for the four considered scenarios: (a) $\mathsf{NF}(8,2)$, (b) $\mathsf{NF}(8,4)$, (c) $\mathsf{FF}(16)$, and (d) $\mathsf{FF}(32)$. The abscissa is the iteration index and the ordinate is the normalized synthesis error $\epsilon$ in~\eqref{eq:SI_cost}, shown on a logarithmic scale.}
\label{fig:gda_4cases}
\end{figure}

\begin{figure}[!tp]
\centering
\includegraphics[width=0.48\textwidth]{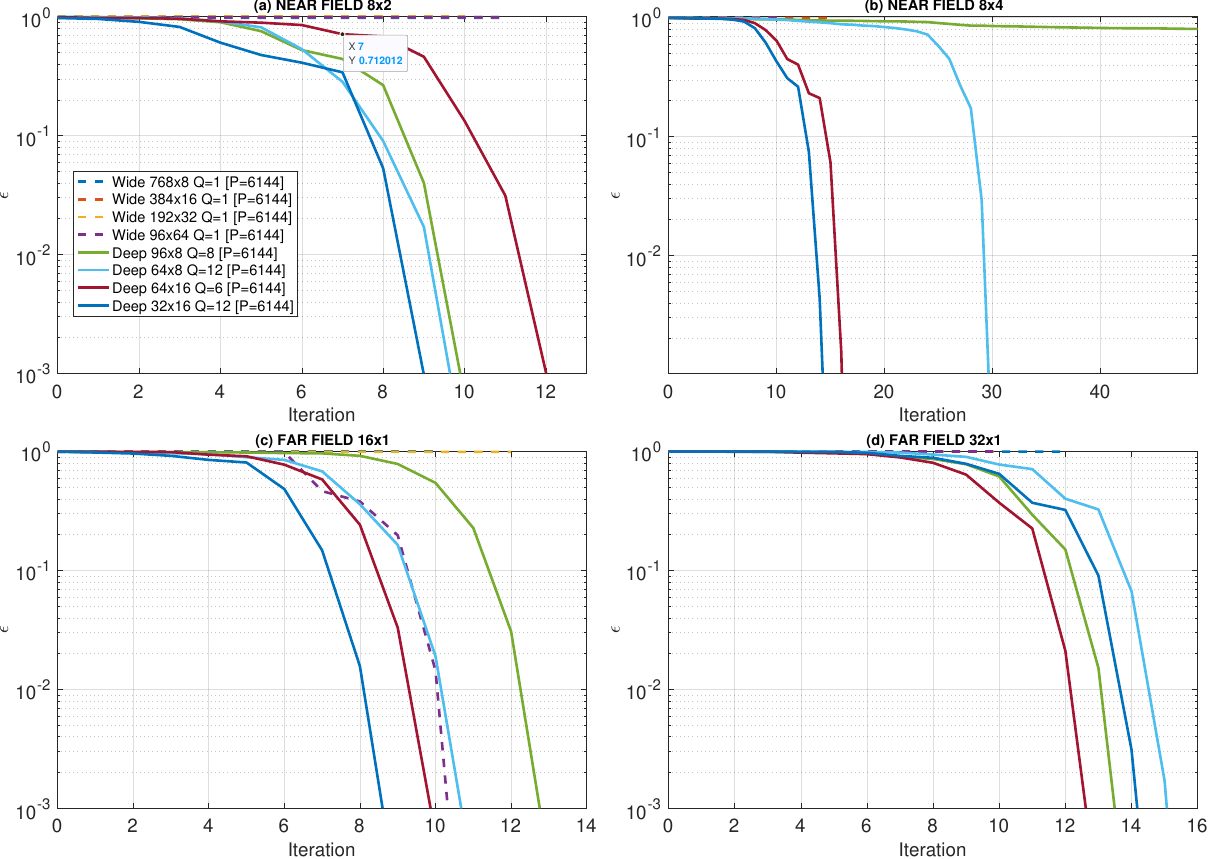}
\caption{Identity-target Gauss--Newton convergence with Armijo backtracking for the four considered scenarios: (a) $\mathsf{NF}(8,2)$, (b) $\mathsf{NF}(8,4)$, (c) $\mathsf{FF}(16)$, and (d) $\mathsf{FF}(32)$. The abscissa is the iteration index and the ordinate is the normalized synthesis error $\epsilon$ in~\eqref{eq:SI_cost}, shown on a logarithmic scale.}
\label{fig:gn_4cases}
\end{figure}

Figures~\ref{fig:gda_4cases} and~\ref{fig:gn_4cases} report the four-scenario comparison for gradient descent and Gauss--Newton, respectively, both with Armijo backtracking and the scale-invariant error in~\eqref{eq:SI_cost}. They qualitatively confirm the advantage of distributing the phase budget across layers observed with reachable random targets. Gauss-Newton generally reaches lower errors in fewer iterations than gradient descent, although iteration counts alone do not measure computational cost. The outcome also depends on geometry and aperture shape: the wide $96\times64$ architecture reaches the displayed accuracy floor with Gauss-Newton in $\mathsf{FF}(16)$, while some deep configurations display slower convergence in NF. These curves provide an additional task-oriented check of the conditioning interpretation.

\section{Conclusions}

This work investigated the role of depth in SIM-based linear transformation synthesis at a fixed phase-control budget. The analysis shows that practical synthesis depends not only on the rank of the transformation Jacobian, but also on the strength and balance of its singular directions. The cascade model provides a physical interpretation of layer complementarity, while direct tests on the multiport Jacobian demonstrate how different layers compensate weak sensitivity directions in the presence of electromagnetic coupling and multiple scattering.

In the considered near- and far-field sensing scenarios, distributing the available controls across multiple layers improves weak-direction sensitivity and synthesis performance. Random-phase Jacobian evaluations establish this architectural advantage independently of the target and optimizer, while experiments with reachable random targets connect it to smaller Gauss--Newton corrections and less restrictive step reductions. These findings identify layer complementarity and Jacobian conditioning as useful criteria for designing SIM architectures under a fixed control budget.

Future work will investigate more closely how the structural properties of inter-layer propagation shape the Jacobian spectrum, with the aim of translating these insights into design guidelines for layer dimensions, spacing, and the allocation of phase controls across the stack.

\bibliographystyle{IEEEtran}
\bibliography{biblio_SIM_v8_conservativa}
\end{document}